\documentclass[aps,prd,reprint,superscriptaddress,amssymb,amsmath]{revtex4-1}

\usepackage{txfonts} 

\usepackage{graphicx}
\usepackage{dcolumn}
\usepackage{bm}
\usepackage{hyperref}

\usepackage{siunitx}

\begin{document}

\title{Magnetic effects on the low-$T/|W|$ instability in differentially rotating neutron stars}

\author{Curran D. Muhlberger}
\email{curran@astro.cornell.edu}
\affiliation{Center for Radiophysics and Space Research, Cornell University, Ithaca, New York 14853, USA}

\author{Fatemeh \surname{Hossein Nouri}}
\affiliation{Department of Physics and Astronomy, Washington State University, Pullman, Washington 99164, USA}

\author{Matthew D. Duez}
\affiliation{Department of Physics and Astronomy, Washington State University, Pullman, Washington 99164, USA}

\author{Francois Foucart}
\affiliation{Canadian Institute for Theoretical Astrophysics, University of Toronto, Toronto, Ontario M5S 3H8, Canada}

\author{Lawrence E. Kidder}
\affiliation{Center for Radiophysics and Space Research, Cornell University, Ithaca, New York 14853, USA}

\author{Christian D. Ott}
\affiliation{Theoretical Astrophysics 350-17, California Institute of Technology, Pasadena, California 91125, USA}

\author{Mark A. Scheel}
\affiliation{Theoretical Astrophysics 350-17, California Institute of Technology, Pasadena, California 91125, USA}

\author{B\'{e}la Szil\'{a}gyi}
\affiliation{Theoretical Astrophysics 350-17, California Institute of Technology, Pasadena, California 91125, USA}

\author{Saul A. Teukolsky}
\affiliation{Center for Radiophysics and Space Research, Cornell University, Ithaca, New York 14853, USA}

\date{\today}

\begin{abstract}
Dynamical instabilities in protoneutron stars may produce gravitational waves
whose observation could shed light on the physics of core-collapse supernovae.
When born with sufficient differential rotation, these stars are susceptible to
a shear instability (the ``low-$T/|W|$ instability''), but such rotation can also amplify magnetic fields to
strengths where they have a considerable impact on the dynamics of the stellar
matter.
Using a new magnetohydrodynamics module for the Spectral Einstein Code,
we have simulated a differentially-rotating neutron star in full 3D to study the effects of magnetic fields on this instability.
Though strong toroidal fields were predicted to suppress the
low-$T/|W|$ instability, we find that they do so only in a small range of field
strengths.
Below \SI{4e13}{G}, poloidal seed fields do not wind up fast enough to have an
effect before the instability saturates, while above \SI{5e14}{G}, magnetic
instabilities can actually amplify a global quadrupole mode (this threshold may be even lower in reality, as small-scale magnetic instabilities remain difficult to resolve numerically).
Thus, the prospects for observing gravitational waves from such systems are not in fact diminished over most of the magnetic parameter space.

Additionally, we report that the detailed development of the low-$T/|W|$
instability, including its growth rate, depends strongly on the particular
numerical methods used.
The high-order methods we employ suggest that growth might be considerably
slower than found in some previous simulations.
\end{abstract}

\maketitle

\section{\label{sec:intro}Introduction}
Stellar core collapse, accretion-induced white dwarf collapse, and
binary neutron star merger all naturally produce rapidly spinning
neutron stars with strong differential rotation.  The resulting neutron stars
could be subject to well-known dynamical instabilities, and the resulting
stellar deformations could produce a strong gravitational wave signal which,
if detected, would provide invaluable information on these violent phenomena.

Global $m=2$ instabilities (perturbations with an azimuthal dependence
of $e^{i m \phi}$) are particularly relevant for gravitational wave
production.  One source of such modes is the dynamical bar
mode instability.  However, this instability only sets in for extremely
high values of the ratio of the rotational kinetic energy $T$
to the gravitational potential energy $W$:  $T/|W|\ge 0.27$ (with small
variations depending on the equation of state and ratio of mass to radius~\cite{1969efe..book.....C,
1998ApJ...497..370T,doi:10.1086/309525,2001ApJ...548..919S}).  Simulations
have revealed another dynamical nonaxisymmetric instability that can
appear at much lower $T/|W|$ if sufficient differential rotation is
present~\cite{doi:10.1086/319634, doi:10.1046/j.1365-8711.2002.05724.x,
2003MNRAS.343..619S, 2003ApJ...595..352S, 2005ApJ...625L.119O, doi:10.1086/507597,
2006MNRAS.368.1429S, 2007CoPhC.177..288C, doi:10.1103/PhysRevLett.98.261101, doi:10.1051/0004-6361:20078577, doi:10.1088/0264-9381/27/11/114104, doi:10.1088/0067-0049/191/2/439}. 
Watts, Andersson, and Jones~\cite{2005ApJ...618L..37W} have given compelling
arguments for identifying this ``low-$T/|W|$ instability'', as it was called, 
as a form of corotation shear instability, similar in basic principle to the
better-known Papaloizou-Pringle instability in thick accretion
disks~\cite{1984MNRAS.208..721P}.  Namely, nonaxisymmetric modes trapped
in a resonant cavity make multiple passes across a corotation radius (the
radius where the mode pattern speed matches the local fluid angular speed)
and are amplified on each pass.  A local minimum of the radial vortensity
profile has been suggested as the mechanism for mode
trapping~\cite{doi:10.1086/507597}.  Simulations of protoneutron stars
indicate that realistic core collapse scenarios can produce stars subject
to this instability~\cite{doi:10.1088/0264-9381/24/12/S10}.
Indeed, the gravitational waves from this instability have been proposed as a distinctive signal from hypothesized magnetorotationally-driven galactic supernovae with rapidly rotating cores~\cite{doi:10.1088/0264-9381/26/6/063001}.

Magnetohydrodynamic simulations have shown
that the dynamical bar mode instability can be suppressed by
magnetic forces, although only for unrealistically high magnetic
field strengths~\cite{doi:10.1088/0004-637X/707/2/1610,doi:10.1103/PhysRevD.88.104028}. 
Fu \& Lai have investigated the effect of a toroidal magnetic field
on the low-$T/|W|$ instability using an analytic model, treating
the star as an infinite cylinder with no vertical
structure~\cite{doi:10.1111/j.1365-2966.2011.18296.x}.  Because of the
strong differential rotation, a more modest poloidal
seed field ($\sim \SI{e14}{G}$) could wind up to a sufficiently strong toroidal field
($\sim \SI{e16}{G}$) within the growth time of the instability (around \SI{30}{ms}).  The
protoneutron stars most likely subject to the low-$T/|W|$ instability have
strong differential rotation and potential for magnetorotational dynamo action,
and in such stars magnetic fields of this magnitude are
plausible~\cite{2009A&A...498..241O}.  Magnetic suppression could therefore
eliminate the potential gravitational wave signal of core-collapse
supernovae.  However, Fu \& Lai's model makes a number of strong simplifying
assumptions:  cylindrical stars, a
polytropic equation of state, and purely toroidal fields.  These could lead
to the neglect of other important magnetohydrodynamical effects and instabilities. 
Thus, simulations of more realistic configurations in full 3D are needed to
evaluate the robustness of the suppression mechanism.

In this work, we simulate the effects of magnetic fields on
differentially-rotating neutron stars susceptible to the low-$T/|W|$
instability, and we do so using a new magnetohydrodynamics (MHD)
module for the Spectral Einstein Code (\textsc{SpEC})\footnote{\url{http://www.black-holes.org/SpEC.html}}.
The instability is indeed suppressed for a narrow range of strong
seed magnetic fields, but the more commonly observed behavior is for either
magnetic fields to be too weak to affect the global quadrupole mode or for them
to be sufficiently strong for magnetic instabilities to set in and actually amplify
the mode.  In general, we find gravitational waves comparable in magnitude
to the unmagnetized case.

\subsection{Notation}
Physical equations in this work are written in geometrized units where the speed
of light $c$ and the gravitational constant $G$ are set equal to 1.
Residual dimensions can be expressed as powers of mass, for which we choose the mass of the Sun, $M_\odot$, as the unit.
When discussing electromagnetic fields in the context of our simulation
formalism and stability analysis, we adopt the Lorentz-Heaviside convention,
absorbing a factor of $1/\sqrt{4\pi}$ into the definition of the magnetic field
$\bm{B}$.
However, when presenting physical results, we express all quantities in
CGS-Gaussian units.
In particular, $\bm{B}_\text{LH} = \bm{B}_\text{G} / \sqrt{4\pi}$.

We denote the Cartesian coordinates of space by $x$, $y$, $z$.
The coordinate distance from the origin of our system is denoted by $r \equiv \sqrt{x^2 + y^2 + z^2}$.
When cylindrical coordinates are used, $\varpi \equiv \sqrt{x^2 + y^2}$ represents the coordinate distance to the $z$-axis, and $\phi \equiv \tan^{-1}(y/x)$ defines a point's azimuthal angle.

Tensor indices from the beginning of the Latin alphabet ($a$, $b$, \ldots) represent spacetime components without reference to any particular coordinate system, while indices from the Greek alphabet ($\mu$, $\nu$, \ldots) range from $0$ to $3$ and correspond to components in our Cartesian coordinate system of $(t, x, y, z)$.  Indices from the middle of the Latin alphabet ($i$, $j$, \ldots) range from $1$ to $3$ and represent spatial Cartesian components.

\section{\label{sec:methods}Numerical methods}
To simulate the behavior of magnetized, differentially-rotating neutron stars, we solve Einstein's equations of general relativity coupled to both the relativistic Euler equations for a perfect fluid and the induction equation of ideal MHD.
The solution is found using \textsc{SpEC}, which implements a hybrid of spectral and finite volume methods~\cite{doi:10.1103/PhysRevD.78.104015}.
As in previous studies conducted with this code, the spacetime metric and its derivatives are evolved on a multidomain pseudospectral grid, while the hydrodynamic variables are restricted to a uniform rectilinear grid encompassing all of the matter in the system and are evolved in conservative form using a high-resolution shock-capturing finite volume scheme.
This work introduces the magnetic field as a new degree of freedom and treats its evolution with an upwind constrained transport scheme on a staggered grid.

The details of our numerical treatment of this system of equations are described in Appendix~\ref{sec:numerical-methods}.
Here we present our definitions for quantities used throughout the rest of the work:

The spacetime metric $g_{ab}$ is decomposed into $3+1$ form with 3-metric $\gamma_{ij}$, lapse $\alpha$, and shift vector $\beta^i$ (see, e.g., Baumgarte \& Shapiro~\cite{isbn:9780521514071}).
The determinant of the 3-metric is denoted by $\gamma$.
The matter in the system is modeled as a perfect fluid with rest-mass density $\rho$, specific internal energy $\epsilon$, and 4-velocity $u^a$.
An equation of state relates $\rho$ and $\epsilon$ to the fluid's pressure $P$,
and from these, the relativistic specific enthalpy is $h = 1 + \epsilon + P/\rho$.
We denote the Lorentz factor corresponding to the fluid's velocity by $W_L\equiv \alpha u^t$.

To this we add an electromagnetic field with Faraday tensor $F^{ab}$, from which we define the magnetic field in a spatial slice to be $B^i = \alpha (\star F^{0i})$ (where $\star F^{\mu \nu}$ is the Hodge dual of the Faraday tensor).
Several quantities of interest are naturally expressed in terms of $b^a$, the magnetic field in a frame co-moving with the fluid:
\begin{equation}
b^a = (\star F^{a b}) u_b \,.
\end{equation}
We adopt the assumptions of ideal MHD; namely, that the fluid is perfectly conducting.

\section{Setup}

\subsection{Physical system}
Since our purpose is to study the effect of magnetic field strength and
configuration on the low-$T/|W|$ instability, we focus here on one system
that, in the unmagnetized case, is subject to this instability.
We choose one of the differentially rotating neutron star
models studied by Corvino et al.~\cite{doi:10.1088/0264-9381/27/11/114104},
namely their configuration \texttt{M.1.200}, which they indeed find to
be unstable.
The star has a baryon mass of $M_b = \SI{2.44}{M_{$\odot$}}$, a central density
of $\rho_c = \SI{1.16e-3}{M_{$\odot$}^{-2}}$, and a ratio of kinetic
to gravitational potential energy of $T/|W|=0.2$ (low enough to
avoid the high-$T/|W|$ dynamical bar mode instability, which becomes accessible for $T/|W|\gtrsim0.24$~\cite{doi:10.1086/309525, doi:10.1103/PhysRevD.75.044023}).
The degenerate component of the equation of state is given by the SLy
model~\cite{doi:10.1051/0004-6361:20011402}, which we implement via
the fitting formula introduced by Shibata et al.~\cite{doi:10.1103/PhysRevD.71.084021}.
Thermal contributions to the pressure and internal energy are included by a
simple $\Gamma$-law addition to the equation of state (see Shibata et al., Duez et al.~\cite{doi:10.1103/PhysRevD.78.104015}), where we have chosen $\Gamma_\text{th}=2$.
At the start of simulations, the temperature of the star is set to zero.
Thus, we ignore for the purposes of this study the
significant thermal energy that would be found in a realistic protoneutron
star or binary post-merger remnant scenario, but we do model
the dominant cold nuclear physics component of the equation of state.

For the initial state of the star, we create an axisymmetric nonmagnetized
equilibrium solution of the Einstein equations.
Differential rotation is a key requirement for the instability and is incorporated by setting the initial angular velocity, $\Omega \equiv v^\phi$, according to
\begin{equation}
\begin{split}
\Omega_c - \Omega &= \hat{A}^{-2}u^tu_{\phi} \\
&= \frac{1}{\hat{A}^2 R_e^2}
\left[ \frac{(\Omega - \omega)r^2 \sin^2(\theta)
e^{-2 \nu}}{1 - (\Omega - \omega)^2 r^2 \sin^2(\theta)
e^{-2 \nu}} \right] \,,
\end{split}
\end{equation}
where $R_e$ is the coordinate equatorial radius, $\Omega_c$ is the central
angular velocity, and $\hat{A}$ is a
dimensionless parameter characterizing the strength of differential rotation.
For the initial state of the system under study,
$R_e=\SI{7.8}{M_{$\odot$}}$,
$\Omega_c=2\pi \times \SI{3.0}{kHz}$, and $\hat{A}=1$.  The ratio of polar to equatorial coordinate
radii is $R_p/R_e=0.414$.
We compute the equilibrium configuration using the code of Cook,
Shapiro, and Teukolsky~\cite{doi:10.1086/171849}.

Since the equilibrium data are axisymmetric to numerical precision, we seed the
star with a small $m=2$ perturbation in order to make the initial perturbation
resolution-independent and its subsequent growth numerically convergent.
This perturbation is applied to the rest-mass density and takes the form
\begin{equation}
\rho \rightarrow \rho \left(1 + \delta_2 \frac{x^2 - y^2}{R_e^2} \right) \,.
\end{equation}
The size of the initial perturbation is $\delta_2 = \num{2e-5}$.
This yields an initial distortion [see Eq.~(\ref{eq:eta-plus-def})] of $\eta_+ = \num{4.08e-6}$.

The properties of the star in its initial state are summarized in
Table~\ref{tab:starprop}.
While the mass is considerably higher than would be
expected for a protoneutron star (though not implausible for a binary neutron star merger remnant), we expect our conclusions regarding the interaction of magnetic fields and the low-$T/|W|$ instability to apply qualitatively to lower-mass systems.
Several properties differ slightly from those of Corvino et al.'s \texttt{M.1.200}, so while we expect the overall evolution to be quite similar, we should not expect perfect correspondence in quantitative measurements.

\begin{table}
\caption{\label{tab:starprop}Basic properties of the neutron star.
$R_e$ is the equatorial coordinate radius, and $R_p$ is the polar coordinate radius.
$\Delta \Omega$ is the angular frequency range---the difference between the central and equatorial rotation frequencies.}
\begin{ruledtabular}
\begin{tabular}{llll}
  & $G,c,M_\odot=1$ & cgs \\
\colrule
$M_0$ & 2.44 & \SI{4.85e33}{g} \\
$M_\text{ADM}$ & 2.19 & \SI{4.35e33}{g} \\
$R_p/R_e$ & 0.414 & 0.414 \\
$\rho_c$ & 0.00116 & \SI{0.717e15}{g.cm^{-3}} \\
$\Omega_c$ & 0.0922 & $2.98 \times 2 \pi$~kHz \\
$\Delta \Omega$ & 0.0650 & $2.10\times 2 \pi$~kHz \\
\end{tabular}
\end{ruledtabular}
\end{table}

\begin{figure}
\includegraphics[width=\linewidth]{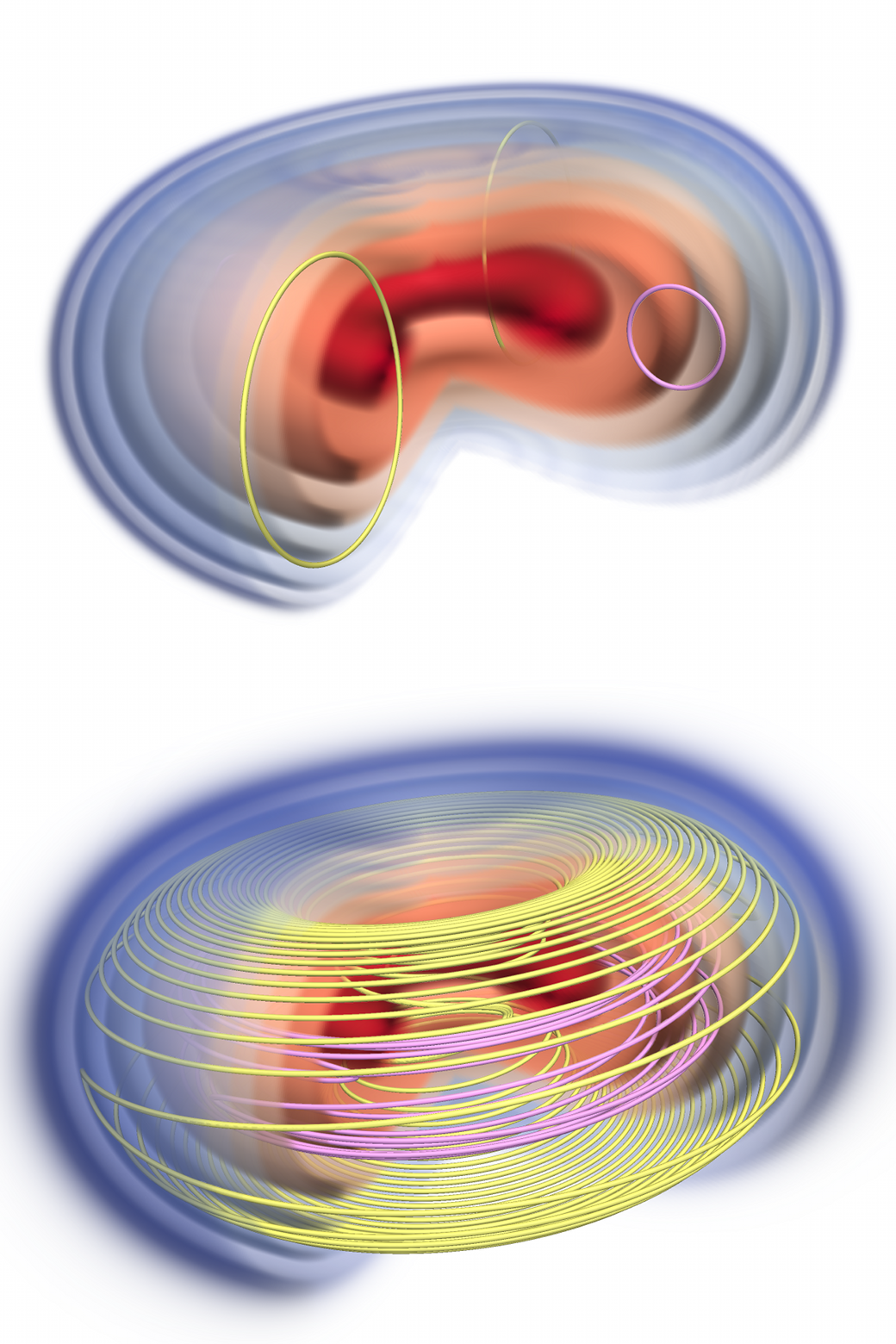}
\caption{\label{fig:vol}(color online).  Illustrations of magnetic field lines at early
($t=0$, above) and intermediate ($t=2160$, below) times.  Contours represent
regions of similar rest-mass density.
Magnetic field lines are seeded at coordinate radii of \SI{2}{M_{$\odot$}} (yellow) and \SI{4}{M_{$\odot$}} (pink).}
\end{figure}

Finally, we introduce a seed poloidal magnetic field.  Following a standard
practice in the numerical literature
(e.g.,~\cite{doi:10.1086/375866,doi:10.1086/422244,doi:10.1103/PhysRevD.83.044014}),
we introduce a toroidal vector potential with strength
\begin{equation}
A_\phi = A_b \varpi^2 \max(P-P_\text{cut}, 0)^{n_s} \,,
\end{equation}
where $A_b$ sets the
overall strength of the resulting $B$-field, $n_s$ controls the smoothness of
the field, and the cutoff pressure $P_\text{cut}$ (set to 4\% of the central pressure) confines the initial field
to regions of high-density matter.
The vector potential is evaluated at cell edges, with a fourth-order curl operator producing the initial $B$-field at cell faces.
This field is then superimposed on top of the unmagnetized equilibrium solution.
While not formally self-consistent, at the field strengths we consider we expect both the deviation from equilibrium and the constraint violations in the equations of general relativity to have negligible effects on our conclusions.
Specifically, the norm of the generalized harmonic constraint energy increased by $<1\%$ with the addition of the magnetic field.
Selected field lines for the initial and evolved states of the star are
illustrated in Fig.~\ref{fig:vol}.

We explored a region of the two-parameter space $A_b \times n_s$.
However, it is more intuitive to talk about magnetic field strengths measured in
Gauss than the poloidal coefficient $A_b$.
The magnetic configurations studied are summarized in Table~\ref{tab:magprops}, which reports both the maximum strength of the $B$-field at $t=0$ as well as a representative initial field strength $B_0$ that more closely reflects the average field in the star.
We assign this representative strength to each magnetic field configuration by measuring the early growth of the magnetic energy within the star, hereafter labeled $H_B$ [see Eq.~(\ref{eq:mag-energy})], and fitting to it the formula
\begin{equation}
H_B \approx B_0^2 \left(\frac{\Delta \Omega^2 R^3}{6}\right) t^2
\end{equation}
to solve for $B_0$.
Here we take $\Delta \Omega = 2.1 \times 2 \pi$~kHz and $R=\SI{15.3}{km}$ (the proper equatorial radius, as opposed to the isotropic coordinate radius reported earlier).
This formula was also used by Fu \& Lai in their
analysis~\cite{doi:10.1111/j.1365-2966.2011.18296.x}, easing comparisons with that work.

The dynamical importance of the magnetic field can be inferred from the ratio
of the gas to magnetic pressure $\beta = 2P/b^2$.
For our strongest initial
field, $\beta$ starts no lower than \num{3.8e2}.

\begin{table}
\caption{\label{tab:magprops}Summary of the magnetic configurations studied.
$B_\text{max}$ is the maximum strength of the initial poloidal magnetic field, $B_0$ is its ``representative'' strength as defined in the text, and $\beta_\text{min}$ is the minimum ratio of fluid pressure to magnetic pressure found initially in the interior of the star.
}
\begin{ruledtabular}
\begin{tabular}{lllll}
  $A_b$ [$G,c,M_\odot=1$] & $n_s$ & $B_\text{max}/\si{G}$ & $B_0/\si{G}$ & $\beta_\text{min}$ \\
\colrule
0 & n/a & 0 & 0 & $\infty$ \\
0.00768 & 1 & \num{2.5e14} & $4\times 10^{13}$ & \num{1.1e6} \\
0.0379 & 1 & \num{1.3e15} & $2\times 10^{14}$ & \num{5.2e4} \\
0.0892 & 1 & \num{2.9e15} & $5\times 10^{14}$ & \num{9.5e3} \\
0.444 & 1 & \num{1.5e16} & $2\times 10^{15}$ & \num{3.8e2} \\
424 & 2 & \num{1.8e15} & $2\times 10^{14}$ & \num{5.9e5} \\
1000 & 2 & \num{4.1e15} & $5\times 10^{14}$ & \num{1.1e5} \\
\end{tabular}
\end{ruledtabular}
\end{table}

\subsection{Simulation parameters}
We used several evolution grids over the course of this investigation, but our final results were achieved on a ``reference'' finite volume grid with $\Delta x = \Delta y = \SI{0.17}{M_{$\odot$}} = \SI{250}{m}$ and $\Delta z = \SI{0.10}{M_{$\odot$}} = \SI{150}{m}$.
Grids employed during the exploratory phase (discussed in Sec.~\ref{sec:reconstructors}) used uniform resolution and are detailed where mentioned.

\begin{figure}
\includegraphics[width=\linewidth]{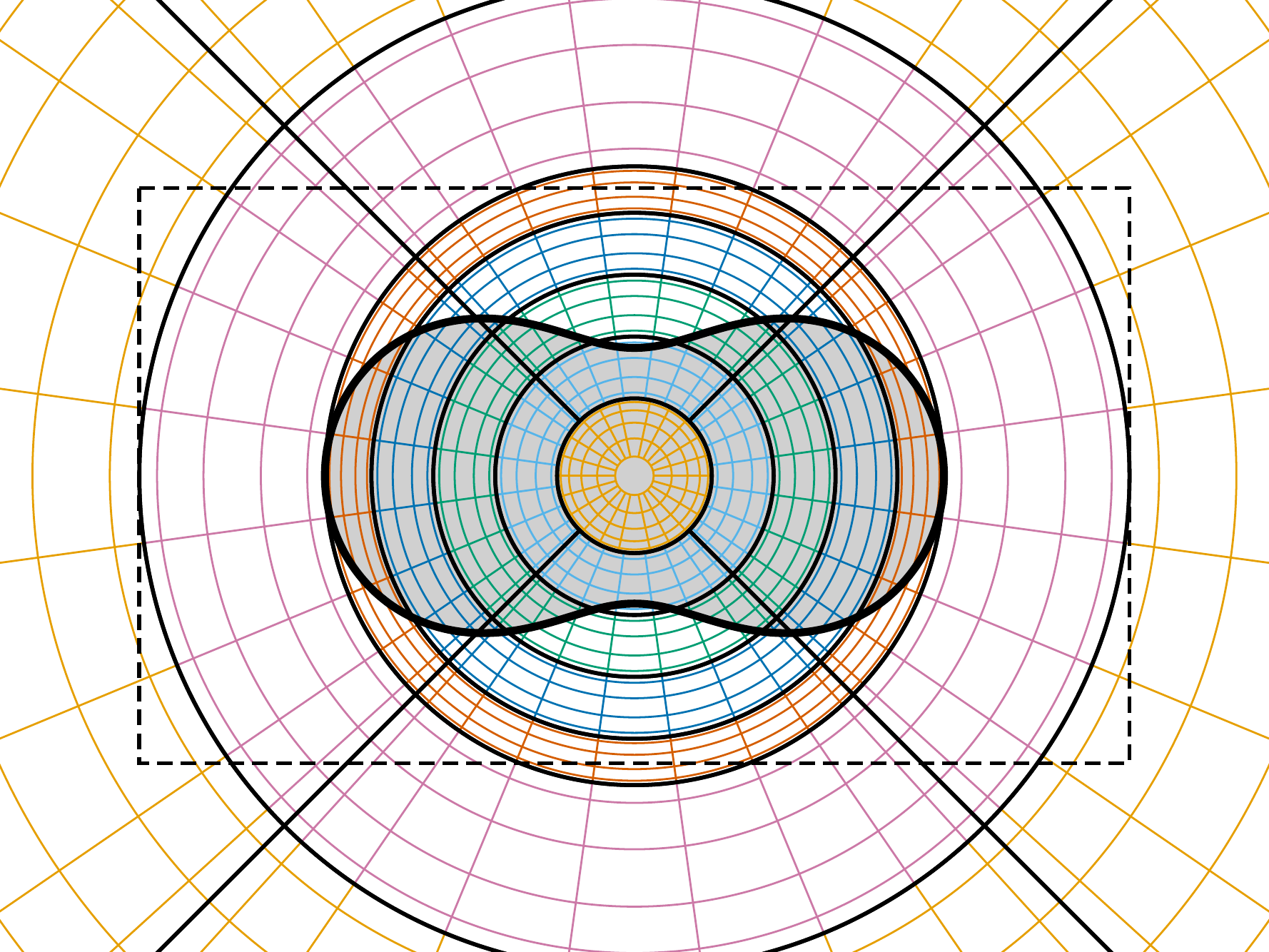}
\caption{\label{fig:ltw-dom}(color online).  Illustration of $x$--$z$ slice of domain decomposition.
The shaded region with a bold outline represents the initial star.
The dashed rectangle represents the finite-difference domain, which has a coordinate width of \SI{25}{M_{$\odot$}} and a coordinate height of \SI{14.5}{M_{$\odot$}}.
For spectral subdomains, the actual reference grid has twice as many collocation points in each direction as are shown in the figure.}
\end{figure}

Our spectral grid (for evolving the spacetime; see Fig.~\ref{fig:ltw-dom}) consists of a filled sphere (using a basis of three-dimensional generalizations of Zernike polynomials; see Appendix~\ref{sec:zernike-spec}) surrounded by layers of ``cubed spheres'' -- products of Chebyshev polynomials distorted to conform to $1/6$ of a spherical shell.
These encompass the entire finite volume grid and are in turn surrounded by true spherical shells (a product of Chebyshev polynomials and spherical harmonics) extending to 300 stellar equatorial radii.
The spectral resolution of our reference grid corresponds to spherical harmonics out to $l=21$ for the central sphere and $l=17$ for the outer spheres.
The radial dimensions of these spheres are resolved by 12 and 11 collocation points, respectively.
The cubed spheres contain 12 radial points and 20 transverse points.

\section{Analysis}
To study the low-$T/|W|$ instability in our simulations and the effects that magnetic fields have on it, we consider several global measures of the simulation results as functions of time.
These include various energy integrals, defined as follows:

Rest mass:
\begin{equation}
M_b = \int \rho W_L \sqrt{\gamma} d^3 x \,.
\end{equation}

Kinetic energy:
\begin{equation}
T = \frac{1}{2} \int \rho h W_L u_i v^i \sqrt{\gamma} d^3 x \,,
\end{equation}
where $v^i \equiv u^i/u^0$.

Internal energy:
\begin{equation}
U = \int \rho W_L \epsilon \sqrt{\gamma} d^3 x \,.
\end{equation}

Magnetic energy:
\begin{equation}
\label{eq:mag-energy}
H_B = \frac{1}{2} \int b^2 W_L \sqrt{\gamma} d^3 x \,.
\end{equation}

Since total energy is conserved (and our hydrodynamic evolution is conservative), we can infer the change in gravitational energy from the sum of the changes in these non-vacuum energies.
Some of this is lost in the form of gravitational waves, which emit \SI{2.1e-4}{M_{$\odot$}} of energy over the duration of the simulation in the unmagnetized case.
Any remaining difference must therefore be a change in the gravitational binding energy of the star.

Following previous studies, we consider the quadrupole moment of the rest mass density about the origin (which is the initial center-of-mass):
\begin{equation}
I^{ij} = \int \rho W_L x^i x^j \sqrt{\gamma} d^3 x \,.
\end{equation}
To reduce this to a scalar measure, we consider two polarizations of the $x$ and $y$ components of the quadrupole tensor,
\begin{align}
\label{eq:eta-plus-def}
\eta_+(t) &\equiv \frac{I^{xx}(t) - I^{yy}(t)}{I^{xx}(0) + I^{yy}(0)} \\
\eta_\times(t) &\equiv \frac{2 I^{xy}(t)}{I^{xx}(0) + I^{yy}(0)} \,,
\end{align}
and, following Corvino et al.~\cite{doi:10.1088/0264-9381/27/11/114104}, take their magnitude to define the ``distortion parameter'' $\eta$:
\begin{equation}
|\eta(t)| = \sqrt{\eta_+^2(t) + \eta_\times^2(t)} \,.
\end{equation}

Note that the numerical atmosphere surrounding the star (see Sec.~\ref{sec:atm}) has the potential to bias integral measurements like those above.
A common solution is to impose density or radius thresholds when summing the integrand.
However, because our fluid grid only covers the region immediately around the star and does not extend into the wave zone, the effect of the atmosphere on these measurements is negligible.

The invariant strength of the magnetic field is simply the magnitude of $b^a$, whose square is equal to
\begin{equation}
\label{eq:mag-vec-sq}
b^2 = \frac{B^2}{W_L^2} + \left[B^i \left(\frac{u^j}{W_L} + \frac{\beta^j}{\alpha}\right) \gamma_{ij}\right]^2 \,.
\end{equation}
To report physical results, we convert this strength to CGS-Gaussian units via
\begin{equation}
\begin{split}
|B_\text{CGS}| &= \frac{\sqrt{4 \pi b^2}}{\SI{1}{M_{$\odot$}}} \left(\frac{c^2}{G M_\odot}\right) \left(\frac{c}{\sqrt{4 \pi \epsilon_0 G}}\right) \times \SI{e4}{G} \\
&= \sqrt{b^2} \times \SI{8.352e19}{G} \,.
\end{split}
\end{equation}

We also consider the evolution of some quantities in a Lagrangian frame of reference.
To do this, we seed ``tracer'' particles in the fluid and evolve their positions according to the fluid velocity in our Eulerian evolution frame.
The resulting trajectories provide useful information in their own right, and observing quantities along those trajectories allows for their Lagrangian analysis.

Finally, in order to accurately monitor the growth of instabilities of arbitrary $m$ in a robust manner, we consider an additional measure of non-axisymmetry that differs from diagnostics used in previous investigations.
Our approach is discussed below.

\subsection{Azimuthal modes}
Previous studies have analyzed the ``Fourier power'' of $m$-modes of a field $\psi$ by integrating the quantity $\psi e^{i m \phi}$.
Some have performed this integral over a ring, capturing the power at a single radius and height within the system~\cite{doi:10.1086/507597, doi:10.1088/0264-9381/24/12/S10}.
Others, including Corvino et al., have performed a volume integral.
While the latter approach incorporates contributions from the entire system, it has several disadvantages.
The integrand is in general discontinuous at the origin for $m>0$, and thus naive numerical computations of $|P_m|$ can produce spurious results (for example, computing a finite volume integral with a gridpoint at the origin will result in non-zero $m>0$ power for axisymmetric data).
Additionally, $m$-modes of $\psi$ whose phase changes with radius or height will be biased (for instance, a tightly wound spiral structure will produce canceling contributions to the integral for each infinitesimal annulus).
Diagnostics defined in terms of multipole moments, like $\eta$, do not suffer the discontinuity problem, but radial cancellations still cause, for instance, the quadrupole moment to be a potentially poor representation for what one would intuitively call ``$m=2$ power.''

A hybrid approach is to sum the power of $\psi$ in several rings, thus sampling the field at multiple heights and radii.
More generally, $\psi$ can be multiplied by a set of orthogonal window functions isolating particular subsets of the domain, with volume integrals used to compute the power of each product.
These functions would approach the origin as $\varpi^m$, ensuring smoothness there, and would be localized at various radii, avoiding cancellation from spiral structure.
A natural choice for such a set of functions are the radial and vertical cardinal functions associated with a basis for functions over a cylinder (for example, the product of Zernike polynomials over a disk with Legendre polynomials in $z$).
These functions are smooth, orthogonal, and generally localized around their corresponding node.

In fact, this approach is equivalent to a spectral measure of $m$-power, defined in Eq.~\ref{eq:mpower}, where the Fourier components of $\psi$ are decomposed into a set of basis functions, and the squared magnitude of the spectral coefficients are summed (see Appendix~\ref{sec:zernike} for proof).
It is this definition of $m$-power, which we denote with $P_m[\psi]$, that we employ in our analysis.
To account for possible center-of-mass motion, the origin is chosen to follow the measured center-of-mass ($\int \bm{x} \rho W_L \sqrt{\gamma} d^3 \bm{x} / \int \rho W_L \sqrt{\gamma} d^3\bm{x}$) of the system.

\section{Results}
Having established the accuracy and convergence of our code on standard test problems (see Appendix~\ref{sec:test-problems}), we can now compare our findings regarding the unmagnetized low-$T/|W|$ instability with previous simulations of the same system, confirming the baseline against which magnetized results will be compared.

\subsection{\label{sec:reconstructors}Unmagnetized instability}
\begin{figure}
\includegraphics[width=\linewidth]{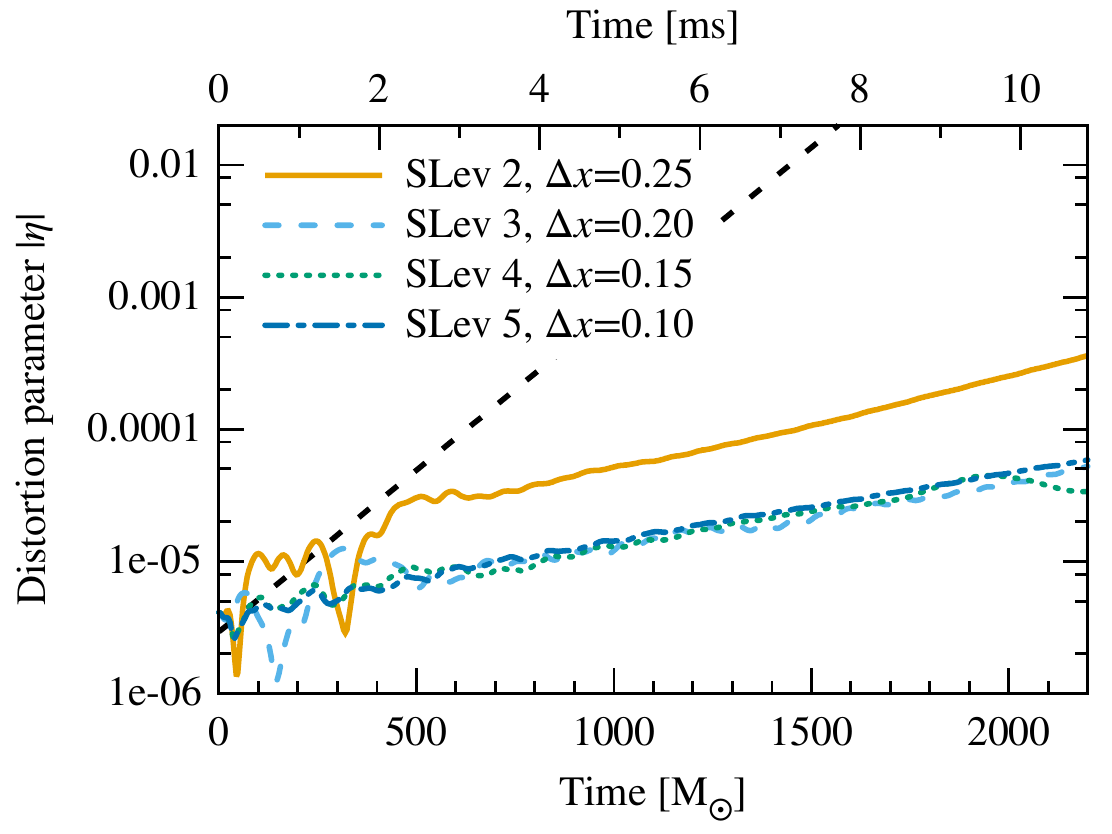}
\caption{\label{fig:weno-consistency}(color online).  Consistency of the growth rate of the
low-$T/|W|$ instability when using WENO5 reconstruction at various resolutions
(no magnetic field is present).
The black dashed line represents the approximate growth rate found by Corvino
et al.\ for \texttt{M.1.200}.
Results from resolutions of
$\Delta x \lesssim \SI{0.2}{M_{$\odot$}}$,
while not formally convergent, are in good agreement and are clearly distinct
from those of Corvino et al.
``SLev'' indicates the spectral resolution level, with higher levels corresponding to finer resolution (the ``reference'' grid uses SLev~4), and grid spacings are measured in solar masses.}
\end{figure}

When simulating the unmagnetized system, we find the behavior of the
low-$T/|W|$ instability to depend sensitively on the reconstruction algorithm
employed by the code (see Appendix~\ref{sec:mhd-impl} for the role and implementation of reconstruction in our evolution scheme).
In particular, the growth of the distortion parameter $|\eta|$ was not
convergent with resolution for the majority of reconstructors considered
(a more thorough investigation is the subject of ongoing work).
We are, however, able to obtain consistent results using WENO5 reconstruction,
as shown in Fig.~\ref{fig:weno-consistency}.

Even when using WENO5 reconstruction, insufficient resolution, particularly in
the vertical direction, can introduce spurious features in the distortion
parameter's evolution at intermediate times and otherwise increase the
simulation's sensitivity to other choices in numerical methods.
We see long-term consistency in the growth of $\eta$ when
$\Delta z \lesssim \SI{0.1}{M_{$\odot$}}$.

\begin{figure}
\includegraphics[width=\linewidth]{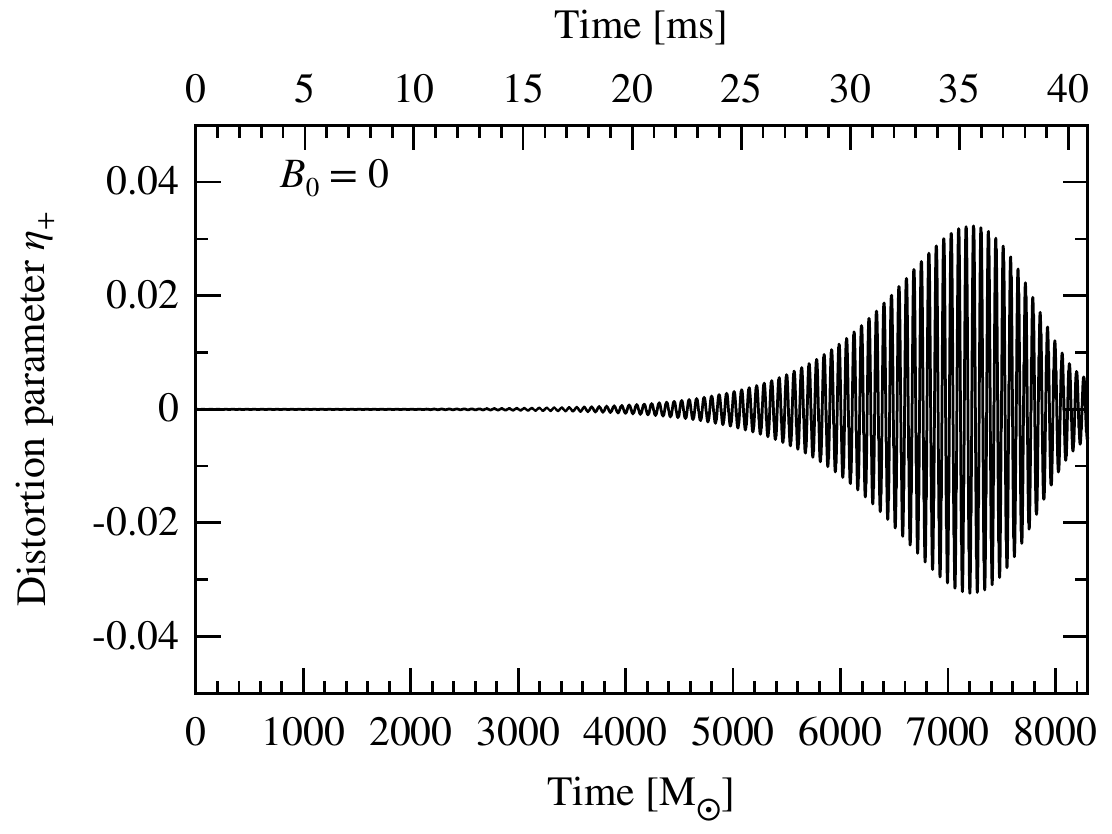}
\caption{\label{fig:ref-eta-lin}Growth and saturation of the unmagnetized
low-$T/|W|$ instability as expressed in the ``plus'' polarization of the distortion parameter $\eta$.
The ``cross'' polarization exhibits the same behavior with a phase shift.
Compare to Corvino et al.\ Fig.~3.}
\end{figure}

We follow the unmagnetized system through the saturation and initial decay of
the instability, as shown in Fig.~\ref{fig:ref-eta-lin}.
The growth is exponential with a time constant of $\tau \approx \SI{3.6}{ms}$,
and the amplitude of the instability saturates when the distortion parameter
reaches $|\eta|_\text{max} \approx 0.035$.
This is the reference against which our magnetized results will be measured.

\begin{figure}
\includegraphics[width=\linewidth]{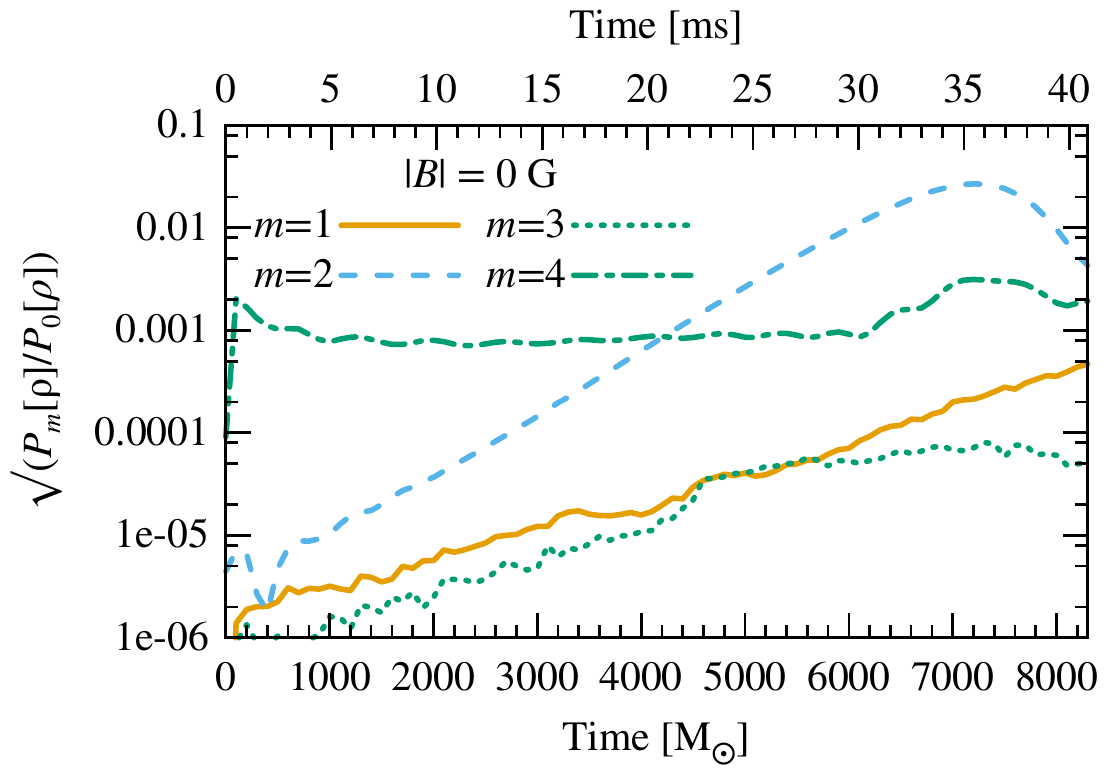}
\caption{\label{fig:mpowerH0}(color online).  Relative power of $\rho$ in azimuthal modes for $m=1$--$4$.
Note that measurements of $m=4$ power have a noise floor of \num{e-3} due to
the Cartesian nature of the grid.}
\end{figure}

Comparing to the results of Corvino et al.~\cite{doi:10.1088/0264-9381/27/11/114104} (who used the piecewise parabolic method for reconstruction), we find a large disagreement in the growth rate of $\eta$.
Our simulations exhibited clean exponential growth for over \SI{30}{ms} with a characteristic time of $\tau \approx \SI{3.6}{ms}$.
For comparison, from Fig.~3 in Corvino et al.'s work we estimate a growth time
of $\tau \approx \SI{0.88}{ms}$.
This rate is illustrated by the dashed line in Fig.~\ref{fig:weno-consistency} and results in saturation of the instability
considerably sooner than in our simulations.
Saturation amplitudes, however, agree to within a factor of two ($0.035$ vs. $0.055$).
Overall, the growth profile we observe for $\eta$ is much more similar to those Corvino et al.\ report for stars with even lower values of $T/|W|$ (0.15 and 0.16), showing smooth exponential growth followed by decay, than what they report for $T/|W|=0.2$.

The relative power of the density perturbation in the lowest few Fourier modes is shown in Fig.~\ref{fig:mpowerH0}.
Unlike Ott et al.~\cite{doi:10.1088/0264-9381/24/12/S10}, but consistent with Scheidegger et al.~\cite{doi:10.1051/0004-6361:20078577} and Corvino et al., we find $m=2$ to be the dominant mode.
This is also the mode whose interaction with magnetic fields was analyzed in detail by Fu \& Lai~\cite{doi:10.1111/j.1365-2966.2011.18296.x}.

\subsection{Magnetic effects}
\begin{figure}
\includegraphics[width=\linewidth]{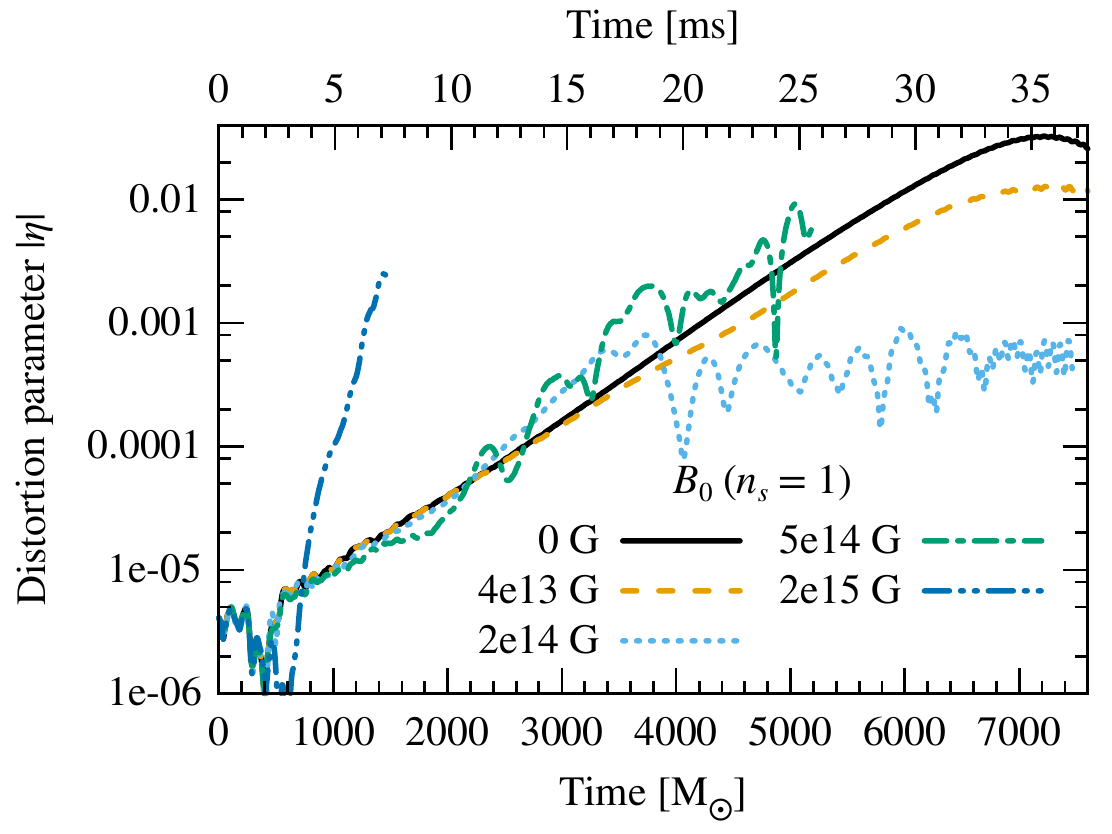}
\caption{\label{fig:eta-ns1}(color online).  Range of behavior of distortion parameter $\eta$ at different magnetic field strengths for $n_s=1$.  Curves that terminate at early times developed significant outflows, making further evolution impractical on our grid.}
\end{figure}

\begin{figure}
\includegraphics[width=\linewidth]{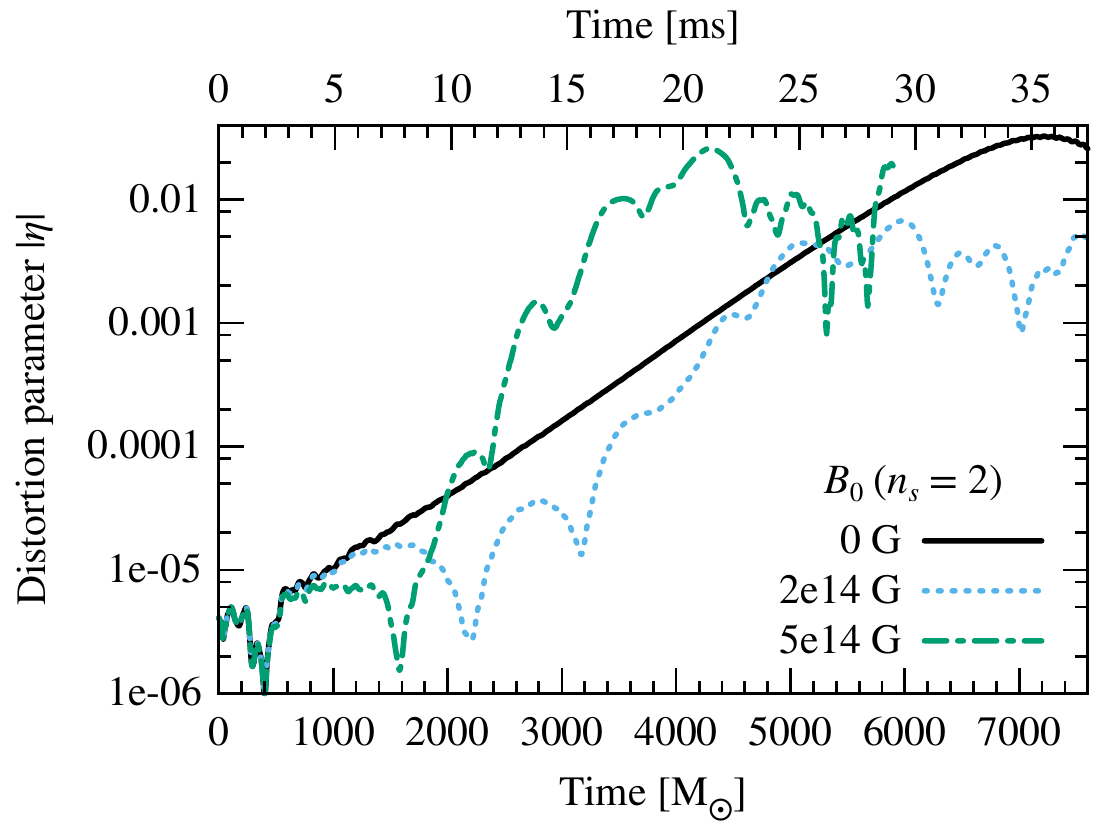}
\caption{\label{fig:eta-ns2}(color online).  Range of behavior of distortion parameter $\eta$ at different magnetic field strengths for $n_s=2$, showing same classes of behavior as when $n_s=1$ (see Fig.~\ref{fig:eta-ns1}).}
\end{figure}

We find that the presence of a magnetic field could have two competing effects
on the growth of the $m=2$ fluid instability.
Simulations with fields of \SI{4e13}{G} and greater demonstrate suppression of the instability, with the distortion parameter saturating at a significantly smaller value ($3$--$50 \times$ lower) than in an unmagnetized star.
Even stronger fields (starting at \SI{5e14}{G}), however, made the star
susceptible to a small-scale (few gridpoints per wavelength) magnetic instability that rapidly amplified the $m=2$ distortion of the star (in addition to other modes).
This instability may operate at lower field strengths as well, but there its effects would not be resolvable at our current resolution.
The net behavior for all simulated cases is plotted in Figs.~\ref{fig:eta-ns1}
and~\ref{fig:eta-ns2} and is qualitatively independent of the seed field
geometry (parameterized by $n_s$; in particular, the threshold for instability
appears to be the same).

Simulations of these magnetically unstable cases were halted prior to the original saturation time, as magnetized outflows of matter began to leave the grid.
Both magnetically-dominated and pressure-dominated matter leave the star relatively isotropically with mildly relativistic velocities ($W_L \lesssim 0.15$).
The stronger the magnetic field, the sooner these outflows develop.
Similar outflows have been noted in previous investigations~\cite{doi:10.1088/2041-8205/785/1/L6, doi:10.1088/2041-8205/785/2/L29}, though due to the small size of our grid, we cannot make quantitative comparisons.

\subsubsection{Suppression of the low-$T/|W|$ instability}

When we observe suppression, we would like to determine whether the mechanism is consistent with that proposed by Fu \& Lai.
Unfortunately, the correspondence is far from clear. 
In particular, while magnetic winding produces peak toroidal field strengths comparable to those considered in their work (and surpassing their threshold for suppression of \SI{2e16}{G}), the total magnetic energy saturates at much lower values than they deem necessary for suppression to take place.
Our runs with initial poloidal field strengths on the order of $B_0 \approx \SI{2e14}{G}$ wind up toroidal fields as strong as \SI{e17}{G} but with magnetic energies of only half a percent of the star's kinetic energy.
For comparison, their model implies that such fields would possess magnetic energy equivalent to 20\% of $T$, which they find is the minimum energy ratio for suppression to occur.

\begin{figure}
\includegraphics[width=\linewidth]{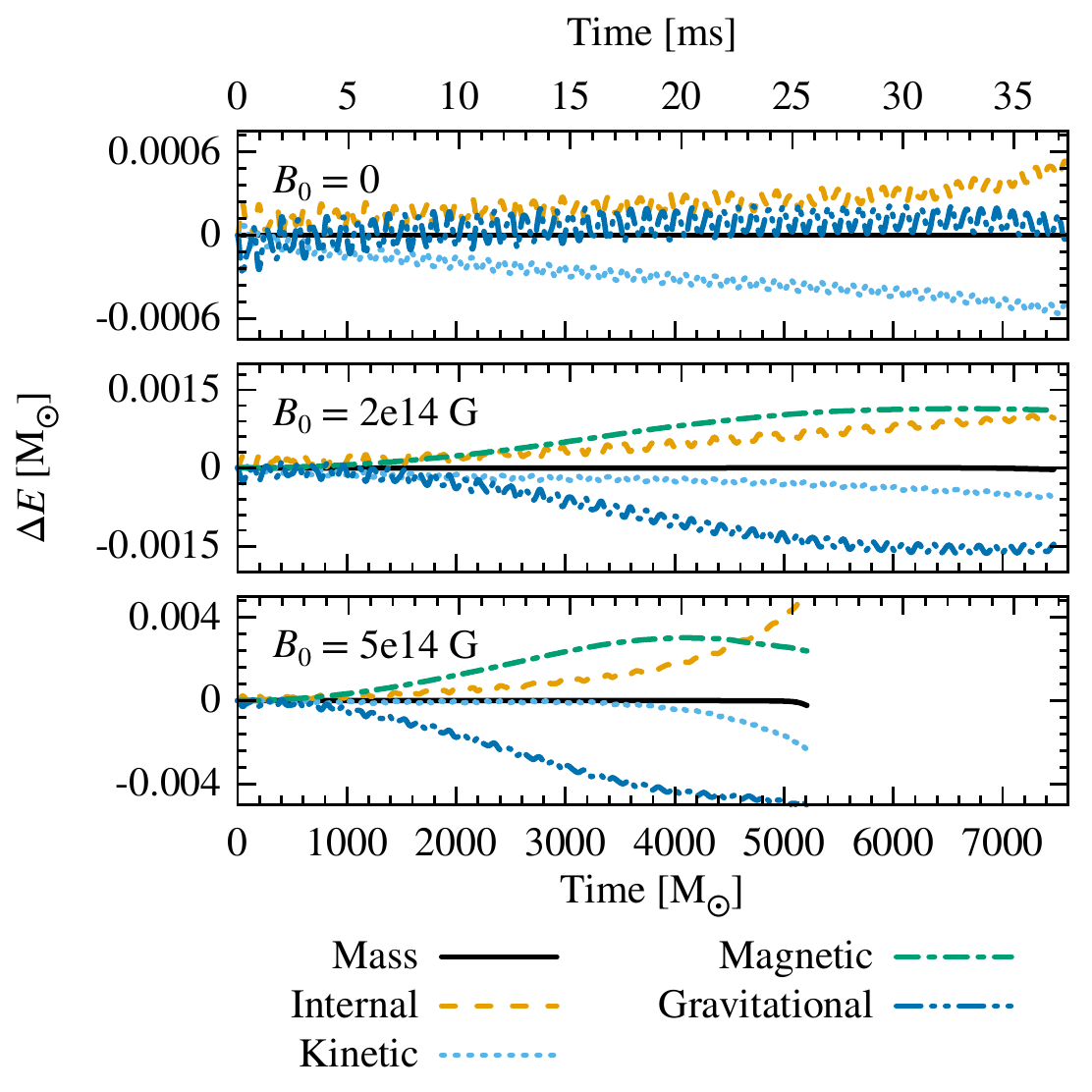}
\caption{\label{fig:energy-exchange}(color online).  Energy exchange for three magnetic field
strengths ($n_s=1$ for each case).
The change in gravitational energy is inferred from the sum of the changes in
the other energies.}
\end{figure}

We see that magnetic winding increases the magnetic energy in the star at the expense of gravitational potential energy, as shown in Fig.~\ref{fig:energy-exchange}, but saturates within \SI{30}{ms} in the cases we considered (prior to the saturation of the low-$T/|W|$ instability).
Matter near the core of the star is compacted, increasing the central density.
The internal energy of the matter also increases in magnetized scenarios, but
the kinetic energy is barely affected in most cases.
For the magnetically-unstable systems, however, kinetic energy from
non-azimuthal fluid velocities grows exponentially at late times as the
rotational kinetic energy begins to decrease at an amplified rate (the separation of rotational and non-rotational kinetic energy is not shown in the figure).
This likely corresponds to small-scale fluid oscillations associated with the
magnetic turbulence described below.

\begin{figure}
\includegraphics[width=\linewidth]{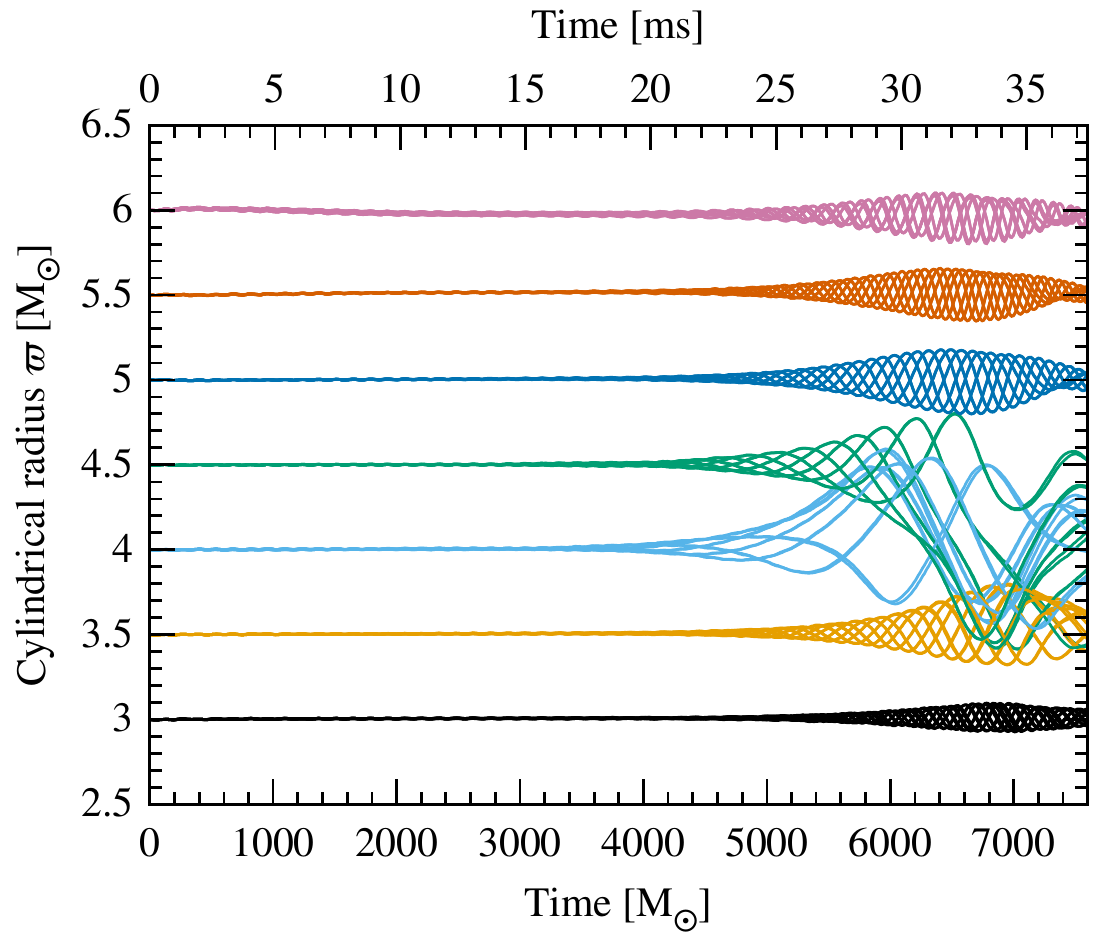}
\caption{\label{fig:tracer-1}(color online).  Lagrangian displacement of tracer particles seeded at various cylindrical radii for an unmagnetized star.  For each initial radius, 12 tracers were distributed uniformly in azimuth.  The corotation radius for this system is at $\varpi \approx \SI{4.25}{M_{$\odot$}}$.}
\end{figure}

Other comparisons are difficult as well.
In Fig.~5 of their paper, Fu \& Lai show that the Lagrangian displacement of fluid elements should diverge at the corotation radius during the low-$T/|W|$ instability, but that this resonance should split in the presence of a strong toroidal magnetic field.
Using tracers, we do see an amplification in radial displacement in the vicinity of the corotation radius in the unmagnetized case (see Fig.~\ref{fig:tracer-1}), but the response is so broad that we cannot resolve any splitting when magnetic fields are added.

\begin{figure*}
\includegraphics[width=\linewidth]{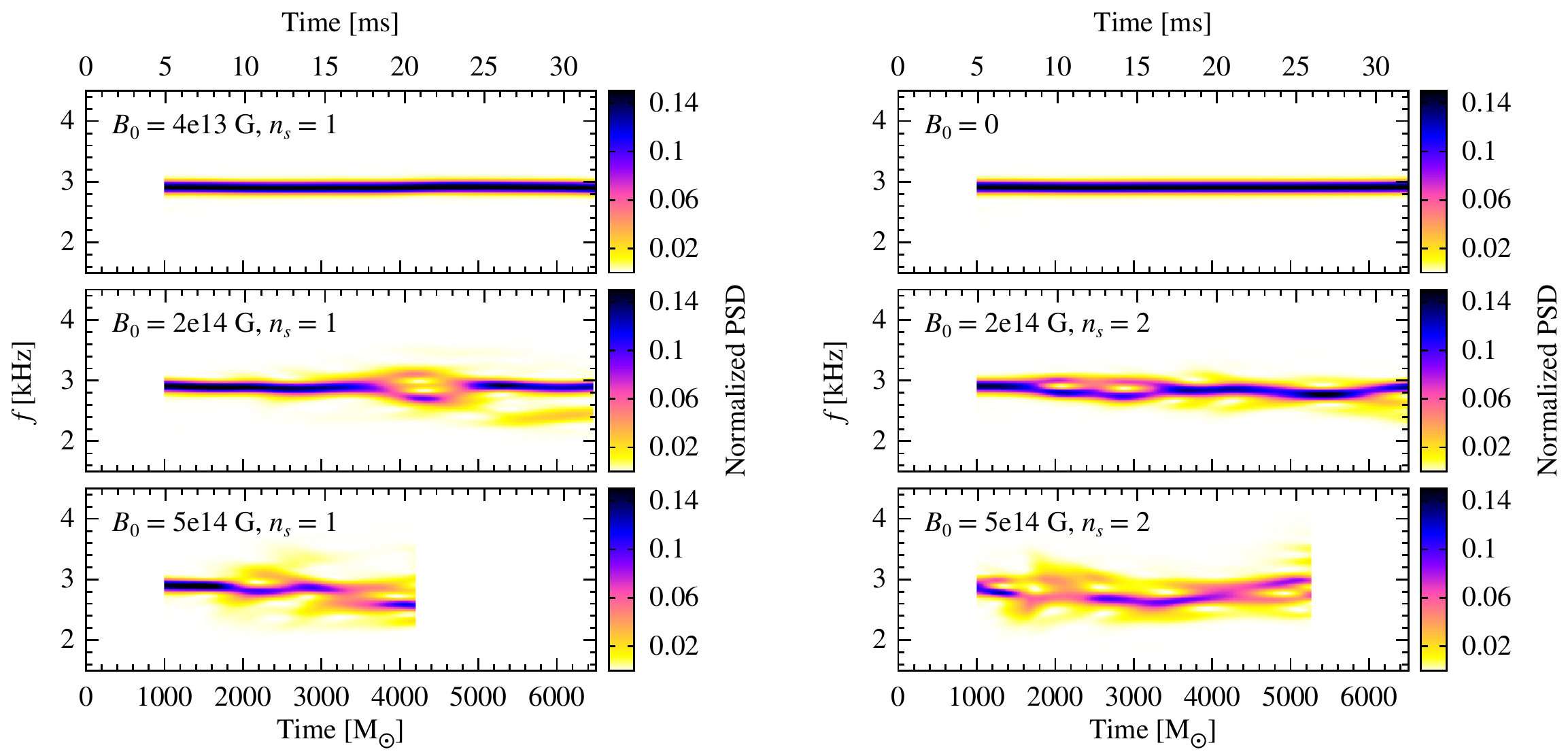}
\caption{\label{fig:spectrograms}(color online).  Spectrograms of the quadrupole moment $I_{xy}$ for six cases.
Power spectral density (PSD) estimated via FFT periodogram using Welch's method with a Hann window.
}
\end{figure*}

Nevertheless, there are clues pointing to a resonance splitting.
In particular, spectrograms of the distortion parameter show a split peak when magnetic suppression is observed (see Fig.~\ref{fig:spectrograms}).
The magnitude of splitting for $B_0=\SI{2e14}{G}$, $n_s=2$ is about $\Delta \omega \approx 2\pi \times \SI{0.1}{kHz}$.
Defining the angular Alfv\'en speed,
\begin{equation}
  \omega_A \equiv B^{\phi}/(\varpi\sqrt{\rho}) \,,
\end{equation}
and the slow magnetosonic wave frequency,
\begin{equation}
  \omega_s \equiv \sqrt{\frac{c_s^2}{c_s^2 + (B^\phi)^2 / \rho}} m \omega_A \,,
\end{equation}
(where $c_s$ is the adiabatic sound speed), resonances are expected at $\Delta \omega = \omega_s$ and (in the full 3D case) $\Delta \omega = m \omega_A$.
In the strongly magnetized regions of the star, the observed splitting agrees with the values of $\omega_s$ and $2 \omega_A$ to within a factor of four.
Given the differences in the particular systems under study, this is reasonably consistent with Fu \& Lai's proposed mechanism.

\subsubsection{Magnetic instability}
When the initial magnetic field exceeds $B_0 = \SI{5e14}{G}$,
our simulations start to exhibit strong magnetic instability.
This instability results both in the amplification of low-$m$ global modes in
the star and in turbulence at the smallest scales we can resolve on our grid.
The marginally-resolved nature of this instability complicates its
identification and interpretation.

\begin{figure}
\includegraphics[width=\linewidth]{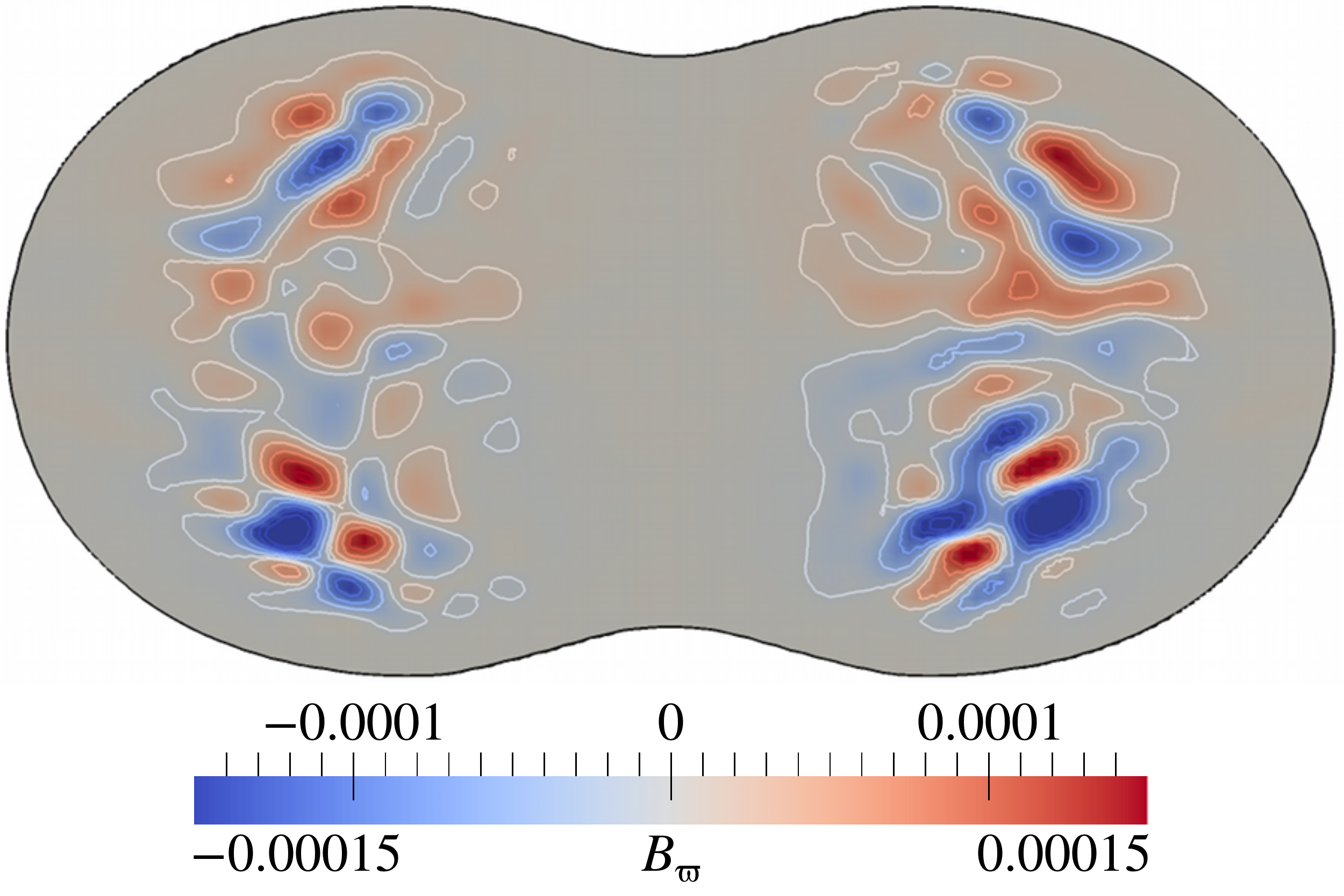}
\caption{\label{fig:bvarpiplot}(color online).  Magnitude of radial component of $B$-field in the
$y$--$z$ plane at $t=\SI{3760}{M_{$\odot$}}$ for $B_0=\SI{5e14}{G}$, $n_s=2$.
}
\end{figure}

\begin{figure}
\includegraphics[width=\linewidth]{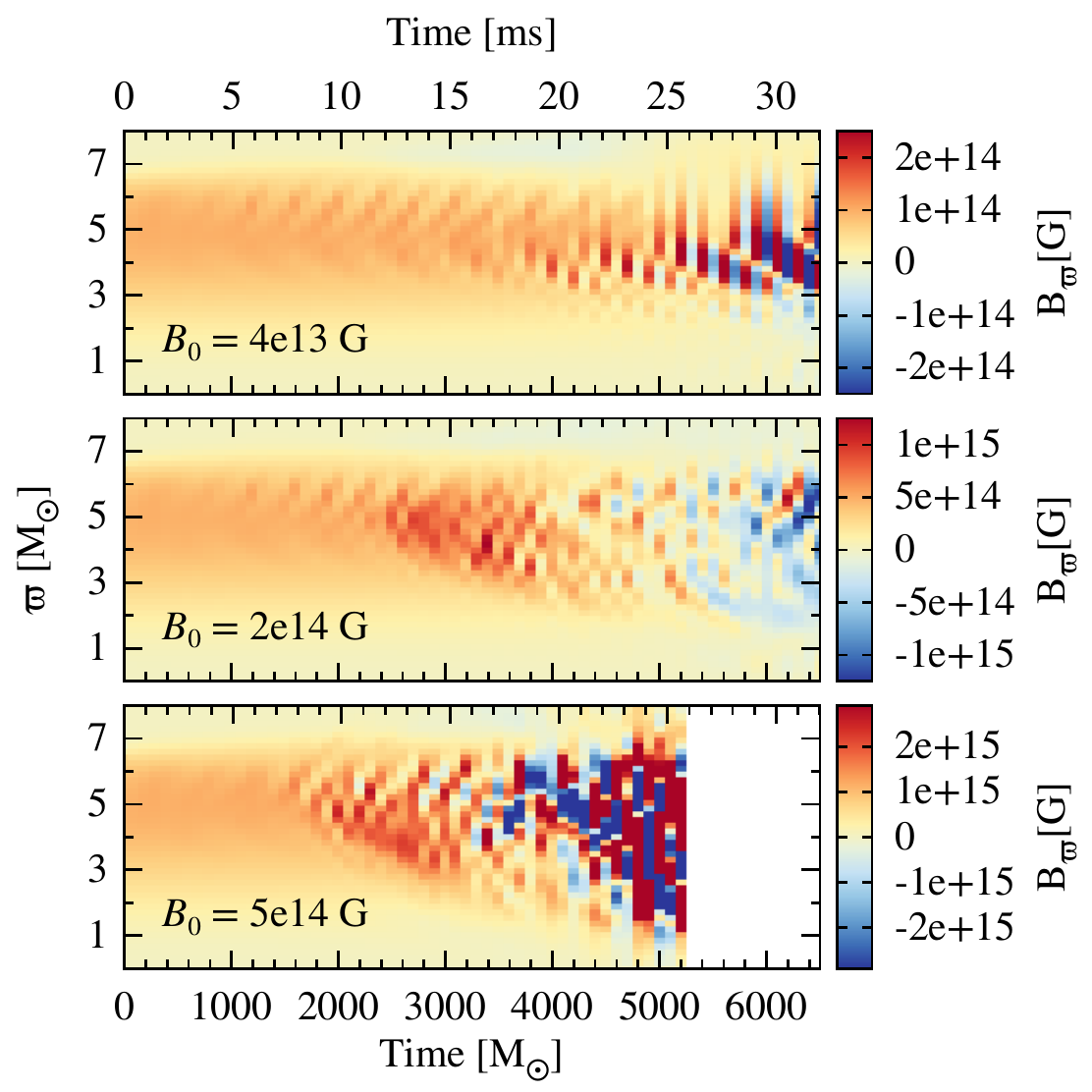}
\caption{\label{fig:turbulence}(color online).  Magnitude of radial component of $B$-field
vs.\ radius vs.\ time in the $z=1$ plane for three configurations ($n_s=1$ in all cases), illustrating
the onset of turbulence.
Colorbars are scaled relative to the initial $B$-field strength.
Plot inspired by the analysis of Franci et al.~\cite{doi:10.1103/PhysRevD.88.104028}.
}
\end{figure}

\begin{figure}
\includegraphics[width=\linewidth]{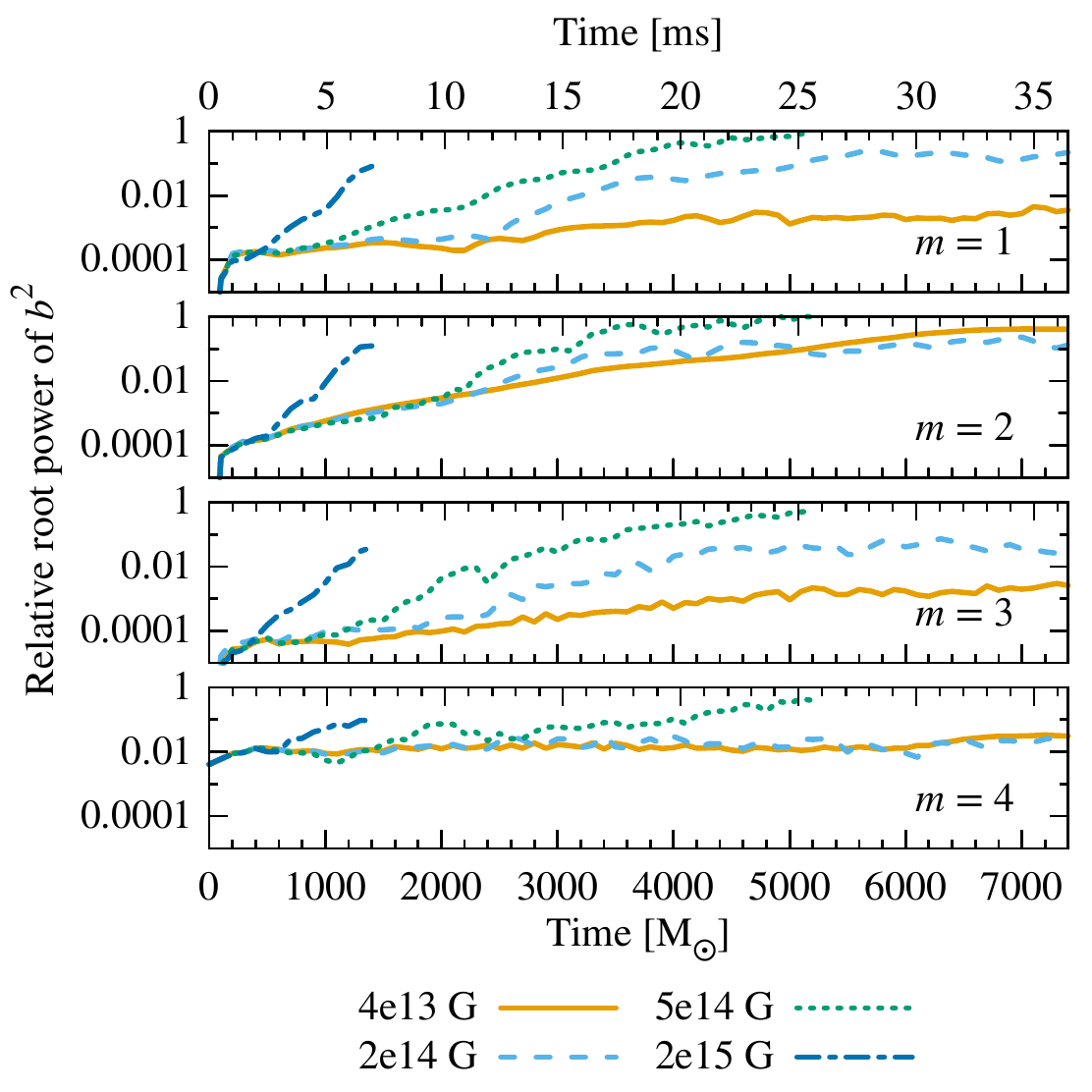}
\caption{\label{fig:mmodes}(color online).  Power of $b^2$ in azimuthal modes for $m=1$--$4$.
Except in the most strongly magnetized systems, the $m=4$ power does not rise above that of the ambient grid mode.
The growing strength of the magnetic field is factored out by normalizing by the $m=0$ power; thus, trends shown here represent growth of the proportional power of nonaxisymmetric modes.}
\end{figure}

The growth of small-scale features is most visible in poloidal field components,
as illustrated in Figs.~\ref{fig:bvarpiplot}~\&~\ref{fig:turbulence}, while
large-scale nonaxisymmetric structure is easily seen in the much stronger
toroidal field (see Fig.~\ref{fig:mmodes}).
The crest-to-crest separation of the poloidal perturbations is measured to be
approximately $\lambda \sim \SI{1}{M_{$\odot$}}$, which is resolved by roughly five gridpoints.
This suggests that the unstable modes are only marginally resolved, so we cannot
expect their subsequent evolution to be more than qualitatively correct (at best).

In fact, magnetically-driven instabilities in the fluid are not unexpected.
Magnetic winding generates a strong toroidal field in the interior of the star,
and toroidal field gradients are potentially unstable to kink (Tayler)
and buoyancy (Parker)
instabilities~\cite{doi:10.1086/148828, ads:1973MNRAS.161..365T,
doi:10.1098/rsta.1978.0066, 1998ApJ...493..291B, arXiv:astro-ph/9907138}.
For a toroidal field centered on the rotation axis, the Tayler instability can
occur at cylindrical radii $\varpi$ less than the radial pressure scale height
$H_P$ (defined as in~\cite{doi:10.1098/rsta.1978.0066,2011A&A...532A..30K}
as $2c_s^2/g_{\varpi}$, with $g_{\varpi}$ denoting the radial acceleration) for positive
$dB^{\phi}/d\varpi$.  Kink instabilities have in fact recently been identified
in 3D magnetized core-collapse simulations~\cite{doi:10.1088/2041-8205/785/2/L29}. 
The Parker instability can be triggered by radial
or vertical field gradients (negative $dB^{\phi}/d\varpi$ for $\varpi>H_P$
or negative $dB^{\phi}/dz$).
The growth rate of the Tayler instability is of order the angular Alfv\'en speed
$\omega_A$ for weak rotation and
$\omega^2_A/\Omega$ for strong rotation, where $\Omega \gg \omega_A$ is the
condition for strong rotation~\cite{ads:1985MNRAS.216..139P}.
Growth timescales for the Parker instability are similar.  Although much
analytic work on field-gradient instabilities assumes weak differential
rotation, the Parker instability has been found
to be operable even in some flows with strong shear~\cite{1994A&A...287..297F}. 
In our magnetically-unstable cases, $\omega_A/\Omega$ is $\mathcal{O}(1/2)$
at the corotation radius, suggesting an intermediate regime between weak and
strong rotation.

In addition to the above-mentioned field gradient-driven instabilities,
differential rotation will also trigger shear-driven instabilities.
The most famous is the classic magnetorotational instability (MRI), an
axisymmetric instability triggered by a nonzero (but arbitrarily small) poloidal
field and an outward-decreasing rotation rate~\cite{doi:10.1086/170270}.
More generally, the MRI can also be found in nonaxisymmetric
configurations~\cite{doi:10.1086/172022, doi:10.1103/RevModPhys.70.1}, in which case the background toroidal field can also
contribute to seeding the instability~\cite{doi:10.1098/rsta.1978.0066,doi:10.1086/172022}.
The fastest-growing unstable mode grows on a timescale of $\sim\Omega^{-1}$ and has
a wavenumber given by
\begin{equation}
\label{eq:MRI}
\Omega/\sqrt{-g_{00}} \sim \bm{k}\cdot\bm{v}_A \approx
\frac{k^\varpi B^\varpi + k^zB^z + mB^{\phi}/\varpi}{\sqrt{\rho h + b^2}}
\end{equation}
(on the relativistic factor, see Siegel et al.~\cite{doi:10.1103/PhysRevD.87.121302}.)  The main challenge for numerical MHD simulations is to resolve the MRI
wavelength $\lambda_\text{MRI}=2\pi|\bm{k}|^{-1}$.
Since the field is usually azimuthally-dominated, we see that $m\neq 0$ modes
are potentially easier to resolve, a fact also recently noted by Franci et
al.~\cite{doi:10.1103/PhysRevD.88.104028}, who resolve MRI-like field growth only in
nonaxisymmetrically-unstable stars.
On the other hand, the growth of a given nonaxisymmetric mode will be expected
to terminate when the mode becomes too tightly wound~\cite{doi:10.1086/172022}.
In fact, it has long been known that even a purely toroidal field can seed a
shear instability~\cite{doi:10.1098/rsta.1978.0066,
doi:10.1086/172022,doi:10.1103/RevModPhys.70.1}, although the growth timescales
tend to be longer than those associated with poloidal seed fields, except for
the case of very high $m$, and in that case even a small poloidal field would be
expected to radically alter the flow~\cite{doi:10.1103/RevModPhys.70.1}.

Given the presence of differential rotation and a poloidal magnetic field, our
system is certainly susceptible to the MRI; what is
less clear is our ability to resolve it.
Siegel et al.~\cite{doi:10.1103/PhysRevD.87.121302} state that a minimum of five
gridpoints per wavelength was required to resolve the MRI in their simulations.
Using Eq.~(\ref{eq:MRI}), we can estimate what the wavelength of the
fastest-growing unstable mode would be at any point in our simulation,
optimizing over propagation directions.
Comparing this to our effective grid resolution in those directions, we find
that when turbulence starts to develop in our systems, there are
$\mathcal{O}(\text{few})$ gridpoints per wavelength in the unstable regions of the star
even for $m=0$ modes, and when considering higher $m$, these unstable regions
begin to meet the criterion of five gridpoints per wavelength.
Therefore, resolving the MRI, if only marginally, is conceivable given our
resolution and magnetic field strengths.

One approach to diagnosing the source of turbulence is to measure the growth
rates of observed instabilities and match them to linear predictions.
As mentioned above, the Tayler and Parker instabilities should grow at a rate
between $\omega_A$ and $\omega_A^2/\Omega$, while the MRI's growth rate is
$\Omega$, independent of the $B$-field magnitude.
The rotational frequency of the star in the region of magnetic instability
(which occurs in the vicinity of the corotation radius) is about
$\Omega \approx 1.45\times 2\pi$~\si{kHz}.

\begin{figure}
\includegraphics[width=\linewidth]{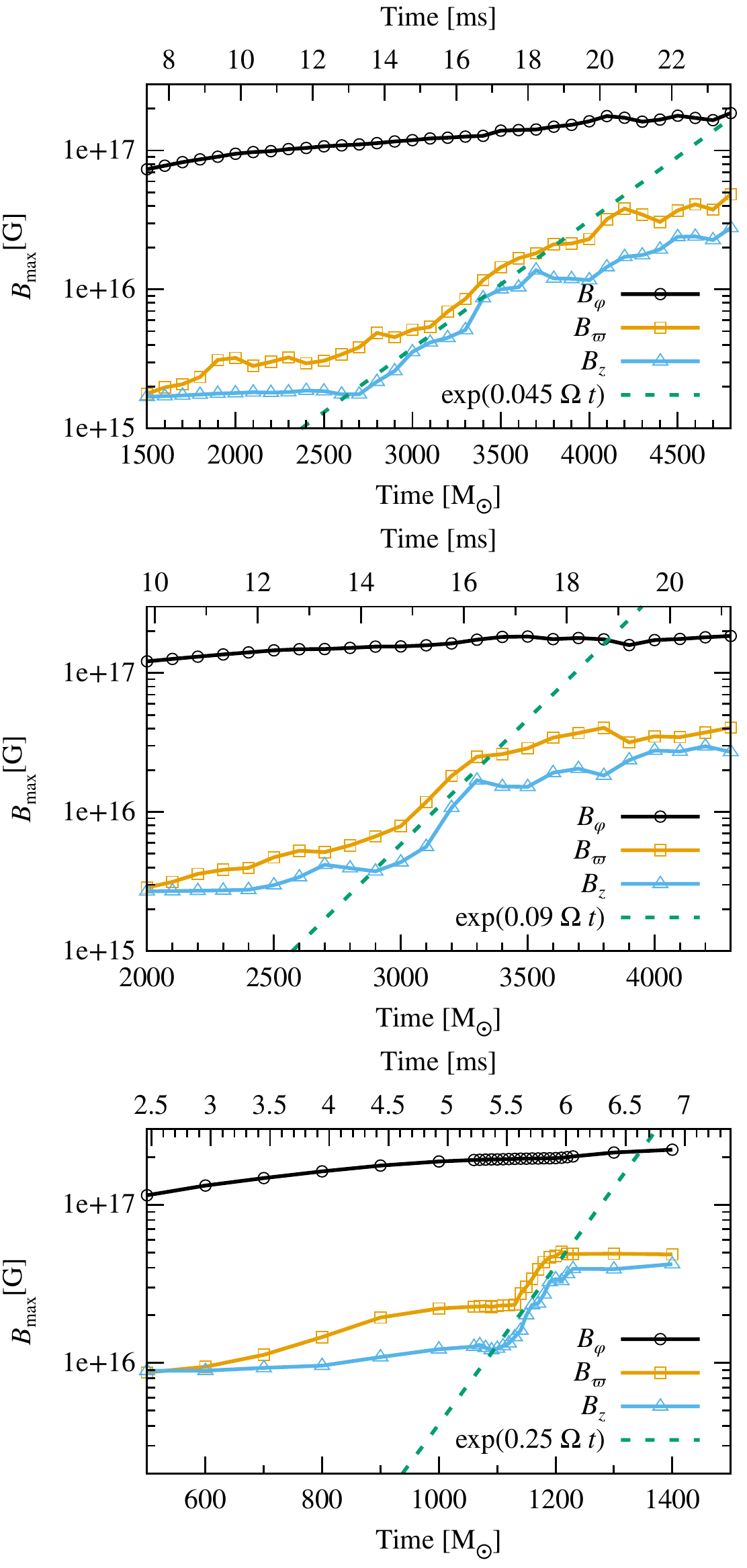}
\caption{\label{fig:growthrates}(color online).  Growth of the maximum of the cylindrical components of the $B$-field for three cases: $B_0=\SI{5e14}{G}$, $n_s=1$ (top), $B_0=\SI{5e14}{G}$, $n_s=2$ (middle), and $B_0=\SI{2e15}{G}$, $n_s=1$ (bottom).
The temporal resolution during the period of rapid growth for the last case is $10\times$ finer than our default.}
\end{figure}

Looking at the growth of the most magnetized point on the grid (see
Fig.~\ref{fig:growthrates}) reveals exponential behavior at rates that increase with the magnetic field strength.
This scaling, in addition to the magnitude of the rates, is incompatible with
the MRI (while the expected rate of $\Omega$ is an approximation derived from
accretion disks, the numerical prefactor for our system is expected to be
$\mathcal{O}(3/4)$, insufficient to explain the discrepancy).

Considering the field gradient-driven instabilities, the ``weak rotation''
rate of $\omega_A$ is too large as well and also does not match the observed
scaling with $B$-field strength.
The ``strong rotation'' prediction, however, while still larger than observed, is only off by a factor of a few and is the closest match to the data in terms of scaling.
This suggests that, while the MRI is potentially resolvable with our techniques,
the observed local maximum $B$-field growth is most attributable
to field gradient instabilities.
Shear instabilities are almost certainly still present and impacting the dynamics, however, and likely play a large role in less-magnetized cases where we currently cannot resolve them.
In fact, their expected growth rates suggest that they would dominate the dynamics on relevant timescales were they resolved.

\subsection{Detectability}
To help put these results in an astrophysical context, we consider the
detectability of gravitational waves produced by the (unmagnetized) low-$T/|W|$
instability for this system.
We follow the procedure outlined by Sutton~\cite{arXiv:1304.0210}.  Given both
polarizations of the gravitational wave strain, $h_+$ and $h_\times$, at some distance from the
source, define the root-sum-square amplitude $h_\text{rss}$ to be
\begin{equation}
h_\text{rss} = \sqrt{\int \left(h_+^2(t) + h_\times^2(t) \right) dt} \,.
\end{equation}
For a narrow-band signal from a rotating system like ours, we expect the emitted
gravitational wave energy $E_\text{GW}$ to be well-approximated by
\begin{equation}
E_\text{GW} \approx \frac{2}{5} \frac{\pi^2 c^3}{G} f_0^2 r^2 h_\text{rss}^2 \,,
\end{equation}
where $f_0$ is the central frequency of the signal.  The effective detection
range $\mathcal{R}_\text{eff}$ for a narrow-band burst signal is given by
\begin{equation}
\mathcal{R}_\text{eff} = \beta \sqrt{\frac{G}{\pi^2 c^3}\frac{E_\text{GW}}{S(f_0) f_0^2 \rho_\text{det}^2}} \,,
\end{equation}
where $S(f)$ is the one-sided noise power spectrum for the target detector,
$\rho_\text{det}$ is the threshold signal-to-noise ratio for detection, and
$\beta$ is a geometrical factor related to the polarization of the waves.
Specializing to rotating sources, this becomes
\begin{equation}
\mathcal{R}_\text{eff} = 0.698 \frac{r h_\text{rss}}{\rho_\text{det}} \sqrt{\frac{2}{5} \frac{1}{S(f_0)}} \,.
\label{eq:Reff}
\end{equation}

We extract gravitational waves from our simulations at a radius of \SI{400}{M_{$\odot$}} using Regge-Wheeler-Zerilli techniques~\cite{doi:10.1088/0264-9381/26/7/075009} and consider the strains $h_+$ and $h_\times$ for an observer above the axis of rotation.
For the unmagnetized star considered in this work, the gravitational wave
frequency is sharply peaked at \SI{2.9}{kHz} (this is slightly lower than the \SI{3.2}{kHz} primary peak observed by Corvino et al.~\cite{doi:10.1088/0264-9381/27/11/114104}).
If we consider only the instability's initial growth through saturation, the total emitted gravitational wave energy is \SI{3.68e50}{erg} (\SI{2.06e-4}{M_{$\odot$}}).
Using the \texttt{ZERO\_DET\_high\_P} noise curve for Advanced LIGO~\cite{LIGO-T0900288-v3} and a signal-to-noise threshold of $\rho_\text{det}=20$, this instability would be detectable out to \SI{92}{kpc}.

The emitted gravitational wave energy is significantly larger
than what was found in core-collapse supernovae simulations~\cite{doi:10.1103/PhysRevLett.98.261101, doi:10.1051/0004-6361:20078577} 
($E_\text{GW} \sim \num{e46}$--\SI{e47}{erg} for a similar simulation length).
However, the difference can easily be understood by noting
that the neutron star considered in this work rotates significantly more rapidly
(with the wave signal peaking at \SI{2.9}{kHz} vs.\ $\sim \SI{0.9}{kHz}$ in the core-collapse results) and is also more massive than protoneutron stars are expected to be.
Since $E_\text{GW}\propto M^2\Omega^6$, this accounts for most of the difference in the emitted gravitational wave energy.
On the other hand, the more slowly rotating neutron stars emit waves at a more favorable frequency, improving their detectability.

The effect of magnetic fields on detectability is difficult to discern from our data, as outflows prevented us from evolving the most highly magnetized systems long enough to see the instability saturate.
For $B_0 = \SI{5e14}{G}$, $n_s=2$, the distortion parameter peaks nearly as high as the saturation value in the unmagnetized case while the frequency spectrum at that time peaks at a slightly lower (and more favorable) value, suggesting that a gravitational wave signal from magnetic instabilities could be just as detectable as that of the unmagnetized low-$T/|W|$ instability.
On the other hand, mildly magnetized cases exhibit a suppressed distortion parameter with an unchanged frequency spectrum.
Using the quadrupole approximation, and the fact that $\mathcal{R}_\text{eff}$ is linear in $h_\text{rss}$, this means that the effective detection range is decreased by factor of $\sim2.4$ for $B_0=\SI{4e13}{G}$, $n_s=1$, and by a factor of $\sim34$ for $B_0=\SI{2e14}{G}$, $n_s=1$, for an observer above the axis of rotation.

\section{Conclusions}
In writing an MHD module for \textsc{SpEC}, we have expanded the range and fidelity of astrophysical systems that can be simulated while still taking advantage of its highly accurate spacetime evolution.
The future scope of this code includes many systems of contemporary interest, including magnetized compact binary coalescence, but here we focus our attention on instabilities in differentially rotating neutron stars.

Of significant relevance to existing literature regarding these stars is the variability in simulated growth rates when using different resolutions and reconstruction methods.
We find qualitative convergence when using high resolution and high-order reconstruction, but these results differ significantly from those of lower-accuracy techniques and of some previous studies.
Further investigation of such instabilities' delicate dependence on simulation methods is warranted.

Regarding the low-$T/|W|$ instability, it is clear that poloidal magnetic fields on the order of \SI{e14}{G} can have a strong effect on the distribution of mass in differentially rotating neutron stars and therefore on their gravitational wave signatures.
However, while suppression of the instability is feasible, it occurs in a small region of parameter space.
$B$-fields strong enough to enable the suppression mechanism are likely also strong enough to trigger magnetic instabilities, accelerating the growth of a mass quadrupole moment rather than suppressing it.

In our simulations, with clean poloidal initial fields, the window between the onsets of magnetic suppression and magnetic instability -- roughly \SI{4e13}{G}--\SI{5e14}{G} -- is rather small, and future runs with increased resolution may lower the upper bound still further.
Therefore, amplification of matter perturbations seems to be the more likely magnetic effect, with peak amplitudes comparable to those in the unmagnetized case.
The spectrum of the gravitational waves, while perhaps possessing more structure, will also remain peaked near the same frequency.
As a result, even with such extreme field strengths, the net effect on burst detectability is likely minor.

Regarding Fu \& Lai's conclusions, we find some disagreement between their predictions for cylindrical stars and our simulations of realistic ones.
In particular, they concluded that suppression would occur once the magnetic energy $H_B$ reached about 20\% of the kinetic energy $T$.
However, the magnetic energy in our simulations peaks at 0.56\% of $T$, yet we still find suppression in some cases.
Despite this, we agree on the minimum strength of the poloidal seed field, roughly \SI{e14}{G}.
Additionally, the frequency spectrum of the instability is consistent with their proposed mechanism for suppression.

Uncertainties in our investigation include the details of the formation of the star and its seed field, as nature will not be nearly as clean as the system we considered.
Additionally, we expect that if the MRI were fully resolved, it would grow on such a short timescale that it would dominate the effects observed here.

Future work to understand the details of the suppression mechanism could investigate the effects of purely toroidal fields, removing the complications of magnetic winding and the MRI.
On the other hand, the impact of the magnetic instabilities could be better understood by increasing resolution and by extending the simulations to observe their saturation behavior.
Additionally, the systematic effects of reconstruction order and grid resolution on the growth rate of this particular instability warrant further investigation.  Lastly, while this paper has limited itself to studying the growth of instabilities,
the later evolution of such stars, after the commencement of
magnetically-driven driven winds, would be a very astrophysically interesting
subject for future numerical modeling.

\begin{acknowledgments}
We extend our thanks to D.\ Lai for inspiring this investigation, to M.\ Boyle for advice on several occasions, and to F.\ H\'ebert for catching errors in the text.
The authors at Cornell gratefully acknowledge support from National Science Foundation (NSF) Grants No.\ PHY-1306125 and No.\ AST-1333129, while the authors at Caltech acknowledge support from NSF Grants No.\ PHY-1068881 and No.\ AST-1333520 and NSF CAREER Award No.\ PHY-1151197.
Authors at both Caltech and Cornell also thank the Sherman Fairchild Foundation for their support.
F.\ Foucart gratefully acknowledges support from the Vincent and Beatrice Tremaine Postdoctoral Fellowship, from the NSERC of Canada, from the Canada Research Chairs Program, and from the Canadian Institute for Theoretical Astrophysics.
Finally, the authors at WSU acknowledge support through NASA Grant No.\ NNX11AC37G and NSF Grant No.\ PHY-1068243.

Some computations were performed on the GPC supercomputer at the SciNet HPC Consortium~\cite{doi:10.1088/1742-6596/256/1/012026}, funded by the Canada Foundation for Innovation under the auspices of Compute Canada, the Government of Ontario, Ontario Research Fund -- Research Excellence, and the University of Toronto.
This work also used the Extreme Science and Engineering Discovery Environment (XSEDE) through allocations No.\ TG-PHY100033 and No.\ PHY990002, supported by NSF Grant No.\ OCI-1053575.
Additionally, this research was performed in part using the Zwicky computer system operated by the Caltech Center for Advanced Computing Research and funded by NSF MRI No.\ PHY-0960291 and the Sherman Fairchild Foundation.
\end{acknowledgments}

\appendix
\section{Numerical methods}
\label{sec:numerical-methods}

\subsection{Metric evolution}
As in previous studies using \textsc{SpEC}, the spacetime is evolved according to Einstein's equations in generalized harmonic form~\cite{doi:10.1088/0264-9381/23/16/S09}, and the coordinates $x^a$ are assumed to obey
\begin{equation}
g_{ab} \nabla^c \nabla_c x^b = H_a
\end{equation}
for some gauge source function $H_a$ (where $\nabla_a$ is the covariant derivative operator associated with $g_{ab}$).
To reduce the equations to first-order form, we evolve the derivatives of the spacetime metric $g_{ab}$, defined as
\begin{align}
\Phi_{iab} &\equiv \partial_i g_{ab} \\
\Pi_{ab} &\equiv -n^c \partial_c g_{ab} \,,
\end{align}
where $n^a$ is the normal to a spacelike slice.
This slicing defines a $3+1$ decomposition of the metric into a 3-metric $\gamma_{ij}$, lapse $\alpha$, and shift vector $\beta^i$ (see, e.g., Baumgarte \& Shapiro~\cite{isbn:9780521514071}), with line element given by:
\begin{equation}
ds^2 = -\alpha^2 dt^2 + \gamma_{ij}(dx^i + \beta^i dt)(dx^j + \beta^j dt) \,.
\end{equation}
The spacetime variables $g_{ab}$, $\Phi_{iab}$, and $\Pi_{ab}$ are evolved according to the principal parts and constraint damping terms in Appendix~A of Foucart et al.~\cite{doi:10.1103/PhysRevD.87.084006} (augmented with the matter and magnetic source terms described below), and the gauge source $H_a$ is evolved according to the ``frozen'' condition in that work.
The damping parameters for the system considered in this work are distributed according to:

\begin{align}
\gamma_0(r) &= \frac{0.1}{M_\text{NS}} f(r) + \frac{0.1}{M_\text{NS}} \,,\\
\gamma_1(r) &= -1 \,,\\
\gamma_2(r) &= \frac{1.5}{M_\text{NS}} f(r) + \frac{0.1}{M_\text{NS}} \,,
\end{align}

where $f(r)$ is given by:
\begin{equation}
f(r) = e^{-r/(6 M_\text{NS})}
\end{equation}
and $M_\text{NS}$ is the ADM mass of the neutron star.

The presence of matter and magnetic fields results in a non-zero stress-energy tensor $T_{ab}$, and this shows up in additional source terms when evolving the spacetime fields.
In particular, the vacuum evolution equation for $\Pi_{ab}$ is modified as follows:
\begin{equation}
\partial_t \Pi_{ab} = \cdots - 2 \alpha \left(T_{ab} - \frac{1}{2} g_{ab} T^{cd} g_{cd} \right) \,.
\end{equation}

The stress-energy tensor for our treatment of MHD is given in Eq.~(\ref{eq:stress-energy}).
Note that we expect the magnetic contributions to $T_{ab}$ to be small, even for our strongest field strengths (magnetic pressure is at most 1\% of fluid pressure at $t=0$).

\subsection{\label{sec:mhd-impl}Magnetohydrodynamics}
The stress-energy tensor of a magnetized perfect fluid, as described in Sec.~\ref{sec:methods}, is given by
\begin{equation}
\label{eq:stress-energy}
T_{ab}=\rho h u_a u_b + P g_{ab} + F_{ac}{F_b}^c - \frac{1}{4}F^{cd}F_{cd} g_{ab} \,.
\end{equation}

Additionally, we adopt the assumption of ideal MHD that the fluid is perfectly conducting:
\begin{equation}
F^{ab}u_b = 0
\end{equation}
(that is, the electric field vanishes in a frame co-moving with the fluid).
This eliminates the electric field as an independent quantity and leaves eight
degrees of freedom: five for the fluid and three for the magnetic field. 

The state of the fluid at each gridpoint is represented in the code by
the ``primitive variables'' $\rho$, $T$, $u_i$, and $B^i$, where $T$ (not to be confused with kinetic energy) is
a variable, related to the temperature, parameterizing the thermal pressure. 
The precise relationship of $T$ to the temperature and thermal pressure is
allowed to vary with the equation of state.  Given $\rho$ and $T$, the
equation of state specifies the pressure $P(\rho,T)$ and specific internal
energy $\epsilon(\rho,T)$.

In order to express the equations of their evolution in conservative form, we
recompose them into the following set of ``conservative'' variables:
\begin{align}
\label{rhostardef}
\rho_* &= \sqrt{\gamma} W_L \rho \\
\tilde{\tau} &= \sqrt{\gamma} \left( W_L\rho (W_L h - 1) - P
+ B^2 - \frac{1}{2}\frac{B^2 + (B^i u_i)^2}{W_L^2}
\right) \\
\label{eq:tilde_S_i}
\tilde{S}_i &= \sqrt{\gamma} \left( W_L \rho h u_i
+ \frac{1}{W_L}\left( B^2 u_i - B^j u_j B^k \gamma_{ik} \right)
\right) \\
\label{bconsdef}
\mathcal{B}^i &= \sqrt{\gamma} B^i
\end{align}
(see also, e.g., \cite{isbn:9780521514071,doi:10.1103/PhysRevD.72.044014}).
Here, $\gamma$ is the determinant of the 3-metric,
$W_L\equiv \alpha u^t$ is the Lorentz factor corresponding to the
fluid's velocity, and $B^2 \equiv B^i B^j \gamma_{ij}$.
These ``conservative'' evolved variables map to the set of ``primitive''
variables through an inversion procedure described in Secs.~\ref{sec:mhd-inversion} \&~\ref{sec:ff-inversion}.

The conservative variables are evolved according to:
\begin{align}
\partial_t \rho_* + \partial_i(\rho_* v^j) &= 0 \,,\\
\partial_t \tilde{\tau} + \partial_i(\alpha^2 \sqrt{\gamma} T_{0i} - \rho_* v^i) &= -\alpha \sqrt{\gamma} T^{\mu \nu} \nabla_\nu n_\mu \,,\\
\partial_t \tilde{S}_i + \partial_i(\alpha \sqrt{\gamma} {T^j}_i) &= \frac{1}{2} \alpha \sqrt{\gamma} T^{\mu \nu} \partial_i g_{\mu \nu} \,,
\end{align}
where $v^i = u^i/u^t$ is the ``transport velocity'' of the fluid.

To compute the behavior of the magnetic field, we define an analog to the electric field,
\begin{equation}
\mathcal{E}_i \equiv -[ijk] v^j \mathcal{B}^k \,,
\label{eq:E_field}
\end{equation}
and then evolve the magnetic field according to
\begin{equation}
\partial_t \mathcal{B}^i = -[ijk]\partial_j \mathcal{E}_k \,,
\end{equation}
where $[ijk]$ is $+1$ for an even permutation of the indices and $-1$ for an odd
permutation.
This evolution is constrained by the zero-monopole criterion,
\begin{equation}
\bm{\nabla}^{(3)} \cdot \bm{B} = \partial_i \mathcal{B}^i = 0
\end{equation}
(where $\bm{\nabla}^{(3)}$ is the covariant derivative operator corresponding to the 3-metric).
In general, a numerical evolution scheme for the magnetic field will not preserve this constraint, so we adopt a constrained transport framework (first used by Yee~\cite{doi:10.1109/TAP.1966.1138693} and later for generally relativistic MHD by Evans \& Hawley~\cite{doi:10.1086/166684}) to do so.

Our constrained transport implementation follows the prescription for ``upwind constrained transport'' proposed by Londrillo \& Del Zanna~\cite{doi:10.1016/j.jcp.2003.09.016} and described in detail by Del Zanna et al.\ as implemented in the ECHO code~\cite{doi:10.1051/0004-6361:20077093}.
In particular, the longitudinal components of $\mathcal{B}^i$ are evolved at cell faces.
This presents a convenient definition of magnetic divergence at cell centers as the second-order divided difference of $\mathcal{B}^i$.
The constrained transport algorithm guarantees that the time derivative of this quantity will be zero to machine precision.
When the $B$-field itself is needed at cell centers, fourth-order polynomial interpolation is used, since discontinuities in the longitudinal direction are forbidden.
Such interpolation is also used when metric quantities are needed at cell faces, as these fields are expected to be smooth.

In order to compute the fluxes of the evolution variables, non-smooth matter quantities must be reconstructed at cell faces and edges.
Our code allows a choice of reconstructors, including a second-order monotonized centered (MC2) limiter~\cite{doi:10.1016/0021-9991(77)90095-X} and a fifth-order weighted essentially non-oscillatory (WENO5) scheme~\cite{doi:10.1006/jcph.1994.1187,doi:10.1006/jcph.1996.0130}\footnote{Instead of adding a fixed $\epsilon=\num{e-6}$ to each smoothness indicator $\beta$, we instead add $\epsilon (1 + \sum_i y_i)$ with $\epsilon=\num{e-17}$.}.
The HLL approximate Riemann solver~\cite{doi:10.1137/1025002} determines a single value for the flux on each interface.
Flux derivatives are computed as second-order divided differences, making our scheme formally second-order accurate (that is, we do not perform the DER operation employed by the ECHO code).
However, higher-order reconstructors, while not affecting the convergence rate, can greatly improve the accuracy of the code (see Sec.~\ref{sec:reconstructors}) at the expense of parallelization efficiency (their larger stencils require additional ghost zones).

In common with other high-resolution shock-capturing codes, \textsc{SpEC} requires procedures for inverting
the relationship between primitive and conservative variables, along with a
prescription for maintaining a tenuous atmosphere around the star.
The addition of a magnetic field necessitates changes to these algorithms, the
details of which we describe below.

\subsubsection{\label{sec:mhd-inversion}Full MHD primitive variable recovery}
We mostly follow the prescription of Noble et al.~\cite{doi:10.1086/500349}
for recovering primitive variables from the evolved conservative variables,
that is the task of numerically inverting
equations~\ref{rhostardef}--\ref{bconsdef}.  We define
\begin{align}
\tilde{S}^2 &= \gamma^{ij} \tilde{S}_i \tilde{S}_j \,,\\
H   &= h(\rho, T) \rho W_L^2 \,,
\end{align}
so that the relations between primitive and conservative variables can be written
as
\begin{gather}
\label{S2con2primeq}
\tilde{S}^2 W_L^2 = \gamma (W_L^2-1)(B^2 + H)^2 - W_L^2 \frac{(\tilde{S}_i B^i)^2 (B^2+2H)}{H^2} \,,\\
\label{taucon2primeq}
-\frac{\rho_* W_L^2 + \tilde{\tau} W_L^2}{\sqrt{\gamma}} = \frac{B^2}{2} + W_L^2\left( \frac{(\tilde{S}_iB^i)^2}{2 \gamma H^2} - B^2 - H + P(\rho,T) \right) \,.
\end{gather}

We solve these equations for $(T,W_L^2)$ using the \texttt{gnewton}
method as implemented by the GNU Scientific
Library~\cite{isbn:9780954612078}, subject to the constraint $W_L^2 \geq
1$.  These equations are more challenging for the root-finding
algorithm than the $B=0$ case, especially in cases where the magnetic
and/or kinetic energy of the fluid is large compared to its rest mass
energy.  When the 2D root-finder for $(T,W_L^2)$ fails, we switch to a
simple 1D bracketing algorithm solving for $H$ ($W_L$ is then considered
as a known function of $H$).

\subsubsection{\label{sec:ff-inversion}Low density force-free primitive variable recovery}
Recovery of the full set of primitive variables can be difficult or
impossible at low-density, magnetically-dominated gridpoints.  Fortunately,
it is also unnecessary.  Our treatment of such points is similar to
that in Ref.~\cite{doi:10.1103/PhysRevD.85.064029}.  For each gridpoint, the code first
attempts to solve the full 2D system for $(T,W_L^2)$.  If a root cannot be
found, it checks that the failing gridpoint is in the force-free regime
by checking the following conditions:
\begin{enumerate}
 \item $\rho W_L/B^2 < 0.001$;\\
 \item $B^{2}>\sqrt{\tilde{S}^2/\gamma}$, which is necessary to have
$B^2>\mathcal{E}^2$; \\
 \item $(\tilde{S}_jB^j)^2 / (B^{2}\rho_{*}^{2})<10$ to prevent
very large velocities along field lines.
\end{enumerate}
If the point satisfies these conditions, then the code attempts a
simpler 1D primitive variable recovery that ignores the internal
energy of the gas.

First, we solve for the 4-velocity:
\begin{equation}
u_{i} =
\frac{W_L}{B^{2}}\left(-\frac{\epsilon_{ijk}(\epsilon^{jlm}\tilde{S}_{l}B_{m})B^{k}}{\sqrt{\gamma}B^{2}+\rho_{*} h W_L}+\frac{(\tilde{S}_{j}B^{j})B_{i}}{W_L \rho_*h}
\right)\,.
\end{equation}
Assuming $T=0$, $h=1$, and using the normalization condition
$W_L^2 = 1 + \gamma^{ij}u_i u_j$, we find
\begin{equation}
\label{eq:W^2}
W_L^{2} =
1+\frac{W_L^{2}}{B^{4}}\left(\frac{\epsilon_{ijk}(\epsilon^{jlm}\tilde{S}_{l}B_{m})B^{k}}{\sqrt{\gamma}B^{2}+\rho_{*}W_L}\right)^{2}+\frac{(\tilde{S}_jB^j)^2}{B^{2}\rho_{*}^{2}}
\,.
\end{equation}
The velocity $\bm{u}$ is composed of a parallel (to the magnetic
field) part and a perpendicular part
$W_L^{2}=1+u_{\parallel}^{2}+u_{\perp}^{2}$, so we have
\begin{align}
\label{eq:u_perp}
u_{\perp}^{2} &=
\frac{W_L^{2}}{B^{4}}\left(\frac{\epsilon_{ijk}(\epsilon^{jlm}\tilde{S}_{l}B_{m})B^{k}}{\sqrt{\gamma}B^{2}+\rho_{*}W_L}\right)^{2}
\,,\\
\label{eq:u_par}
u_{\parallel}^{2} &= \frac{(\tilde{S}_jB^j)^2}{B^{2}\rho_{*}^{2}} \,.
\end{align}
Equation~(\ref{eq:W^2}) is solved for $W_L^{2}$ with a 1D Newton-Raphson
root solver; the other variables can be inferred from the solved $W_L$
and the assumed $T=0$.  For force-free points with very low densities,
or force-free points where we fail to solve Eq.~(\ref{eq:W^2}), we
remove the density-dependent terms in Eq.~(\ref{eq:W^2}) and set
$u_i$ to the drift velocity $(u_{\perp})_i$.  We note that the $h=1$
approximation used above would have to be adjusted when using a
nuclear equation of state in which $h(\rho\rightarrow 0,T\rightarrow 0)$
is slightly less than one (i.e. when the binding energy of nucleons
is taken into account, and the specific internal energy of the fluid
is negative when $\rho\rightarrow 0$).

\subsubsection{\label{sec:atm}Atmosphere treatment}
The methods used for the evolution of relativistic fluids often assume
that $\rho > 0$. In order to avoid numerical problems in regions where
no fluid is present, we have to impose $\rho \geq \rho_\text{floor}$
everywhere.  In this simulation, $\rho_\text{floor}$ is set to
$10^{-14}$ and $\rho_\text{floor}/\rho_\text{max}$ is about $8\times
10^{-12}$. However, numerical errors in the evolution of low-density
fluid can easily lead to values of conservative variables for which
the inversion problem has no solution. We thus need appropriate
prescriptions to:
\begin{itemize}
\item Modify the conservative variables, if necessary, to force them to
correspond to some set of primitive variables;
\item Require the primitive variables (mainly $T$ and $u_a$) in the
low-density region to be physically reasonable.
\end{itemize}
For a given $\rho_*$ and $B^i$, limits to the allowable range of
$\tilde{\tau}$ and $\tilde{S}_i$ come from considering the limit of
zero internal energy ($P=0$, $h=1$).  In this limit, we can
write $\tilde{S}^{2}$ as a function of $W_L^{2}$:
\begin{equation}
\label{eq:tildeS2}
\tilde{S}^{2}
= \frac{\rho_*^{2}\left(W_L+\frac{\sqrt{\gamma}B^{2}}{\rho_*}\right)^2 (W_L^2-1)}
{W_L^2+2\frac{\sqrt{\gamma}}{\rho_*}B^{2} \mu^2 W_L +
\frac{\gamma}{\rho_*^{2}}B^{4} \mu^2} \,,
\end{equation}
where $\mu \equiv B^{i} \tilde{S}_i/\sqrt{B^{2} \tilde{S}^{2}}$.  $W_L$
is given by a fifth-order polynomial equation
\begin{equation}
\label{eq:W}
\begin{split}
0 &= W_L^3 + \left(\frac{\sqrt{\gamma}B^{2}}{\rho_*} - \frac{\tilde{\tau}}{\rho_*} -1\right) W_L^2 \\
&\quad - \frac{\sqrt{\gamma}B^{2}}{2\rho_*}\left( 1 + \frac{\mu^2\left(W_L+\frac{\sqrt{\gamma}B^{2}}{\rho_*}\right)^2 (W_L^2-1)}{W_L^2+2\frac{\sqrt{\gamma}}{\rho_*}B^{2} \mu^2 W_L +\frac{\gamma}{\rho_*^{2}}B^{4} \mu^2}\right) \,,
\end{split}
\end{equation}
This equation has a real solution $W_L \geq 1$ if and only if the
condition $B^{2} \leq 2\tilde{\tau} / \sqrt{\gamma}$ is
satisfied.  Thus, we can ``fix'' our conservative variables
($\tilde{\tau}$ and $\tilde{S}_i$) by imposing:
\begin{align}
\tilde{S}_i &\leq \sqrt{\frac{\tilde{S}^{2}_\text{max}}{(\tilde{S}^{0})^2}} \tilde{S}^{0}_i \,,\\
\label{eq:fixedtau}
\tilde{\tau} &\geq \frac{\sqrt{\gamma}B^{2}}{2} \,,
\end{align}
where $\tilde{S}^{2}_\text{max}$ is the solution to
Eqs.~(\ref{eq:tildeS2}) and~(\ref{eq:W}), and $\tilde{S}^{0}$ is the
value of $\tilde{S}$ before it is ``fixed.''  This recipe to fix
conservative variables is similar to what is introduced
by Etienne et al.~\cite{doi:10.1103/PhysRevD.85.064029}, except that they fix $\tilde{\tau}$ and
$\tilde{S}$ using stricter ``sufficient conditions'' for invertibility
[Eqs.~(A48)--(A50) in their work] for points deep inside their black hole
horizon, while for points elsewhere they only fix $\tilde{\tau}$ using
Eq.~(\ref{eq:fixedtau}).

\subsubsection{Additional adjustments to the low-density evolution}
We also impose several restrictions on the low-density fluid in order
to avoid extreme heating and relativistic speeds in the atmosphere.
This must be done differently in magnetospheric regions than in
nonmagnetic regions, because in the former, the fluid velocity encodes
information about the electric field that should not be sacrificed.

For regions with low $B^2/\rho$, we choose a threshold density
$\rho_\text{atm} > \rho_\text{floor}$, and require that for $\rho <
\rho_\text{atm}$ we have $T=0$ and $u_i=0$.  Additionally, in order to
avoid a sharp transition from the ``live'' evolution to the atmosphere
prescription, we add a smoothing region for $\rho_\text{atm} < \rho <
10\rho_\text{atm}$ where we require $h-1 \leq \kappa (h_\text{max}-1)$
and $u^2 \leq \kappa u^2_\text{max}$, with $\kappa = (\rho -
\rho_\text{atm})/(9\rho_\text{atm})$.  $h_\text{max}$ and
$u^2_\text{max}$ are values larger than the enthalpies and velocities
encountered in the high-density region of the simulation.

On the other hand, for magnetically dominated low-density regions, we
have the same treatment as in weakly magnetic regions for $h$ and for
$u_{i\parallel}$ [the component of the 4-velocity along field lines,
cf. Eq.~(\ref{eq:u_par})], i.e. $u_{\parallel}^2 \leq \kappa
u^2_\text{max}$ for $\rho_\text{atm} < \rho < 10\rho_\text{atm}$, and
$u_{i\parallel}=0$ for $\rho < \rho_\text{atm}$.  The perpendicular
(drift) part of velocity [Eq.~(\ref{eq:u_perp})] can contain, even
for very low densities, physically meaningful information about the
electric field, so it is controlled much more weakly, by imposing the
limit  $u^2_{\perp \text{max}} \leq u^2_{\text{max}}$ for  $\rho <
10\rho_\text{atm}$.

Finally, a half-stencil's worth of points are frozen at atmosphere
levels along all outer boundaries.  This ``boundary condition'' avoids
the complexities of one-sided differencing and has no effect on the
bulk evolution of the matter provided that the grid is large enough
(for the system considered here, magnetic fields are initially confined to high-density regions, and we halt the simulation upon the detection of significant outflows).

\subsection{\label{sec:test-problems}Test problems}
The spacetime and hydrodynamics components of \textsc{SpEC} have been tested previously~\cite{doi:10.1103/PhysRevD.75.024006, doi:10.1103/PhysRevD.78.104015}.
Here, we check the performance of our new MHD module, using a similar test suite as Duez et al.~\cite{doi:10.1103/PhysRevD.72.024028}.
In particular, we study its accuracy and convergence by comparing results to known analytical solutions exhibiting a range of non-trivial behaviors, including shocks and strong gravity.

\subsubsection{One-dimensional relativistic tests}
To test the shock-capturing methods used in \textsc{SpEC}, we evolve a set of one-dimensional problems
first proposed by Komissarov~\cite{doi:10.1046/j.1365-8711.1999.02244.x}. The initial data consist of two homogeneous
states separated by a discontinuity at $x=0$. The initial conditions for each test are listed in Table~\ref{tab:shocktests}.
We integrate the relativistic MHD equations from $t=0$ to $t=t_\text{final}$ (also given in Table~\ref{tab:shocktests}).
The fluid follows a $\Gamma$-law equation of state with $\Gamma=4/3$:
\begin{align}
\label{eq:eos-43-pressure}
P &= \rho^{4/3} + \rho T \,,\\
\label{eq:eos-43-epsilon}
\epsilon &= 3 \frac{P}{\rho}\,,
\end{align}
where we have now defined the code's internal temperature variable $T$ for the
$\Gamma$-law case such that $\rho T$ is the thermal pressure of the fluid.
To facilitate comparisons with previously published results, we use the same resolution as
in Duez et al.~\cite{doi:10.1103/PhysRevD.72.024028}, where the same tests were performed (see Figs.~7--8 and Table~II of that work): our numerical
domain covers the region $x=[-2,2]$, and uses 400 grid points (higher resolution results are also provided to test
the convergence of our code). The tests are performed with both the MC2 reconstructor used by Duez et al.\ and the WENO5 reconstructor that we prefer in most of our simulations.
We use fourth-order Runge-Kutta time stepping, with a Courant factor of $0.5$ ($dt=0.005$), except for
the {\it Fast Shock} problem using WENO5 reconstruction, for which we use a Courant factor of $0.25$ (the evolution is unstable
for a Courant factor of $0.5$, an issue which was also noted by Duez et al.\ when using the third-order piecewise parabolic method for reconstruction).

\begin{table*}
\caption{\label{tab:shocktests}Initial data for the shock tests}
\begin{ruledtabular}
\begin{tabular}{lll}
Test & Initial state for $x<0$ & Initial state for $x>0$ \\
\colrule
Fast shock & $\rho=1$, $P=1$  &  $\rho=25.48$, $P=367.5$  \\
($t_\text{final}=2.5$) & $u_i=(25,0,0)$, $B^i=(20,25.02,0)$ & $u_i=(1.091,0.3923,0)$, $B^i=(20,49,0)$ \\
\colrule
Slow shock & $\rho=1$, $P=10$&  $\rho=3.323$, $P=55.36$ \\
($t_\text{final}=2.0$)  & $u_i=(1.53,0,0)$, $B^i=(10,18.28,0)$  & $u_i=(0.9571,-0.6822,0)$, $B^i=(10,14.49,0)$\\
\colrule
Switch-off & $\rho=0.1$, $P=1$&  $\rho=0.562$, $P=10$ \\
($t_\text{final}=1.0$)  & $u_i=(-2,0,0)$, $B^i=(2,0,0)$  & $u_i=(-0.212,-0.590,0)$, $B^i=(2,4.71,0)$\\
\colrule
Switch-on & $\rho=0.00178$, $P=0.1$&  $\rho=0.01$, $P=1$ \\
($t_\text{final}=2.0$)  & $u_i=(-0.765,-1.386,0)$, $B^i=(1,1.022,0)$  & $u_i=(0,0,0)$, $B^i=(1,0,0)$\\
\colrule
Shock tube 1 & $\rho=1$, $P=1000$&  $\rho=0.1$, $P=1$ \\
($t_\text{final}=1.0$)  & $u_i=(0,0,0)$, $B^i=(1,0,0)$  & $u_i=(0,0,0)$, $B^i=(1,0,0)$\\
\colrule
Shock tube 2 & $\rho=1$, $P=30$&  $\rho=0.1$, $P=1$ \\
($t_\text{final}=1.0$)  & $u_i=(0,0,0)$, $B^i=(0,20,0)$  & $u_i=(0,0,0)$, $B^i=(0,0,0)$\\
\colrule
Collision & $\rho=1$, $P=1$&  $\rho=1$, $P=1$ \\
($t_\text{final}=1.22$)  & $u_i=(5,0,0)$, $B^i=(10,10,0)$  & $u_i=(-5,0,0)$, $B^i=(10,-10,0)$\\
\colrule
Wave & $\rho=1$, $P=1$&  $\rho=1$, $P=1$\\
($t_\text{final}=2.5$)  & $u_i=-0.4133 \cdot (0,\cos{x},\sin{x})$, $B^i=(1,\cos{x},\sin{x})$  & $u_i=-0.4133 \cdot (0,\cos{x},\sin{x})$, $B^i=(1,\cos{x},\sin{x})$\\
\end{tabular}
\end{ruledtabular}
\end{table*}

\begin{figure}
\includegraphics[width=\linewidth]{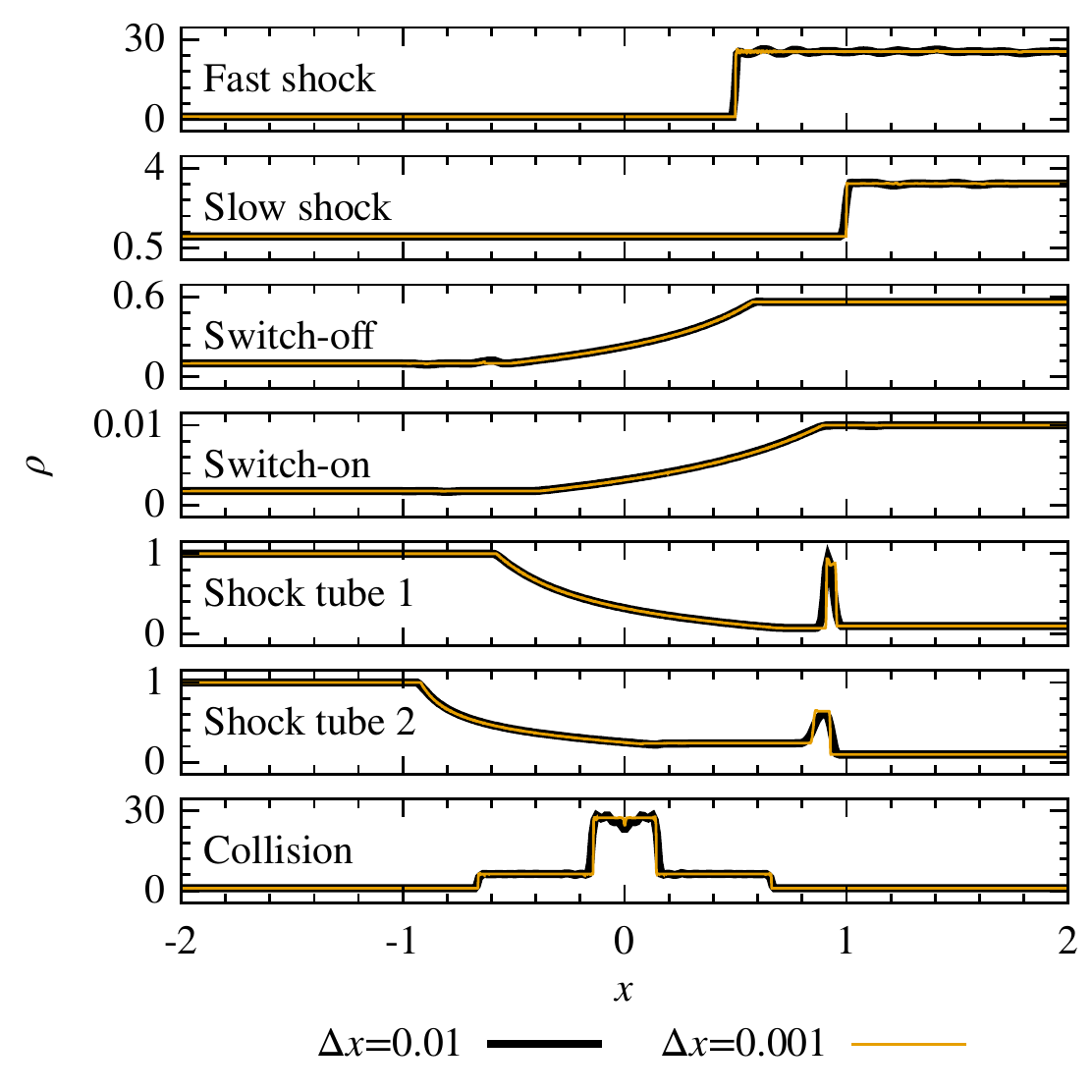}
\caption{\label{fig:shock-rho}(color online).  Rest-mass density at $t=t_\text{final}$ for the shock tests described in Table~\ref{tab:shocktests}, shown for two resolutions ($N=400$ and $N=4000$ points).}
\end{figure}

\begin{figure}
\includegraphics[width=\linewidth]{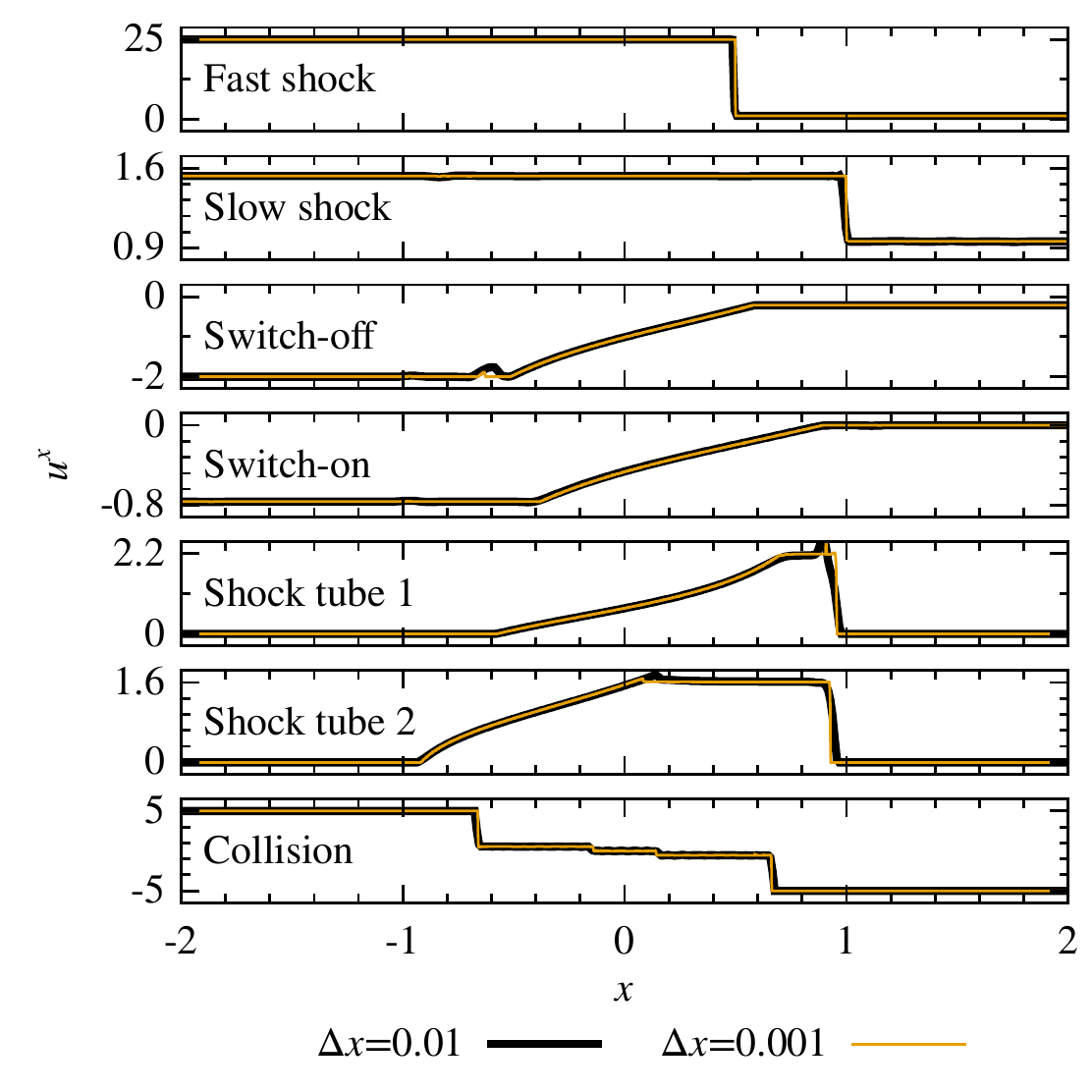}
\caption{\label{fig:shock-ux}(color online).  Velocity at $t=t_\text{final}$ for the shock tests described in Table~\ref{tab:shocktests}, shown for two resolutions ($N=400$ and $N=4000$ points).}
\end{figure}

\begin{figure}
\includegraphics[width=\linewidth]{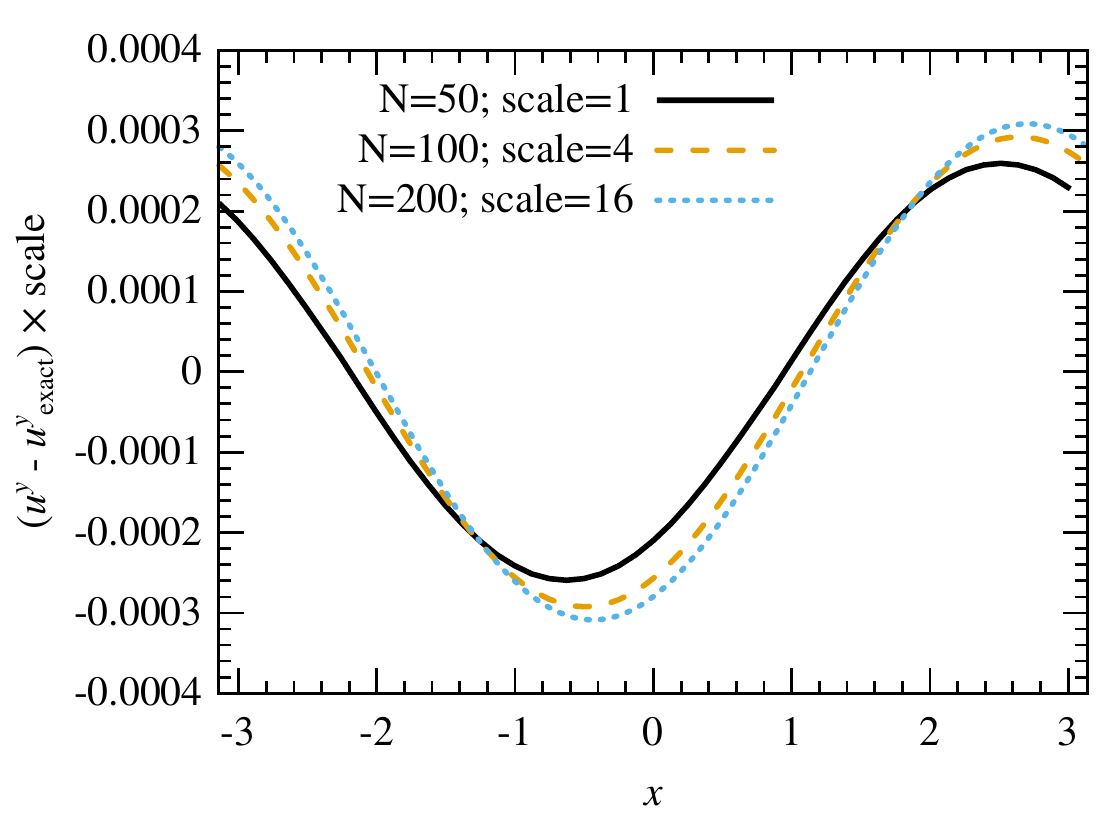}
\caption{\label{fig:wave-uy-err}(color online).  Error in the final value of $u^y$ for the ``wave'' test at 3 resolutions ($N=50$, $N=100$, $N=200$),
rescaled for the expected second-order convergence.}
\end{figure}

\textit{Fast and slow shocks}:
For these two tests, the shock front satisfies the relativistic Rankine-Hugoniot jump conditions~\cite{doi:10.1063/1.522857}.
The exact solution to the evolution of the fluid equation is known, with the shock propagating at constant speed while the fluid variables on each side
of the shock remain constant~\cite{doi:10.1063/1.866479, doi:10.1046/j.1365-8711.1999.02244.x}.
The \textit{fast shock} test is the hardest test for our code: it evolves a strong shock, with the shock front moving relatively slowly on the grid ($0.2c$) but the fluid being highly relativistic (Lorentz factor $W_L=25.02$).
As already noted, it is the only test that is unstable when using a Courant factor of $0.5$ (for WENO5 reconstruction).
It is also fairly sensitive to the choice of variables that are interpolated from cell centers to cell faces when computing the fluxes entering the conservative hydrodynamics equations: if we interpolate the transport velocity $v^i$, the shock evolves as expected, while if we interpolate the spatial components of the 4-velocity $u_i$ the shock immediately stalls.
Considering that in practice, in 3-dimensional evolutions of neutron stars or binary mergers, we do not reliably evolve fluid elements with $W_L\sim 25$ (the occurrence of such high Lorentz factors is prevented by the corrections applied to the velocity and temperature of low-density points in the atmosphere), this difference is unimportant in practice.
The \textit{fast shock} test is mostly evolved in order to verify that our implementation of the MHD equations is correct in the limit of ultra-relativistic fluids.
In fact, because of the practical advantages of using $u_i$ instead of $v^i$, we usually reconstruct the former ($W_L=\sqrt{1+g^{ij}u_iu_j}$ is always well-defined while $W_L=1/\sqrt{1-g_{ij}v^iv^j}$ is not if numerical errors in the low-density regions cause $v^i$ to satisfy $g_{ij}v^iv^j>1$).
In Figs.~\ref{fig:shock-rho} and~\ref{fig:shock-ux}, we show the result of that test when using the MC2 reconstruction method (and reconstructing $v^i$), for 400 and 4000 grid points.
The results converge towards the solution at the expected first-order rate.
The \textit{slow shock} test is generally less extreme.
As in previous studies~\cite{doi:10.1046/j.1365-8711.1999.02244.x,doi:10.1086/374594,doi:10.1103/PhysRevD.72.024028}, we observe that the evolution is very accurate on the left side of the shock, while oscillations are visible on the right side of the shock (see Fig.~\ref{fig:shock-rho}).
Although these oscillations converge away as we increase the resolution, they do so more slowly than expected past 200-400 points in the evolution domain (convergence order of $\sim 0.6$).
This is the only test for which we do not observe at least first-order convergence.

\textit{Other shock tests}: The five other one-dimensional shock tests, for which results are presented in Figs.~\ref{fig:shock-rho} and~\ref{fig:shock-ux},
are comparable to previously published results in accuracy (for the simulations using 400 points),
and convergent when the resolution is increased to 4000 points. As expected, the convergence is fairly
slow (first-order), which explains why sharp features remain visible even at high resolution. These tests
cover a wide range of potential behaviors (shock waves, rarefaction waves, contact discontinuities), and indicate
that the shock capturing methods implemented in \textsc{SpEC} are capable of handling the discontinuities which are likely to
arise in our simulations.

\textit{Wave}: The last one-dimensional test to which we submit our code is the propagation of a wave on
a periodic grid. In this case, all variables are continuous, and the error in the simulations should be
second-order convergent. In the exact solution, the initial profile (given in Table~\ref{tab:shocktests}) simply propagates
with velocity $v=0.3820$. The error in the density $\rho$ at the end of the simulation for 3 different resolutions (50,100 and 200 points per
wavelength) is shown in Fig.~\ref{fig:wave-uy-err},
rescaled for the assumed second-order convergence. Our results also appear in good agreement
with the theoretical predictions for this smooth configuration.

\subsubsection{Bondi accretion}
We also test the ability of our code to evolve a magnetized fluid in the strong gravitational field of a black hole.
We check its ability to maintain stationary and spherically symmetric accretion onto a Schwarzchild black hole according to the relativistic Bondi accretion solution.
This test is nontrivial since we have an extremely strong gravitational field and relativistic fluid which contains nonzero magnetic terms.
There is also an exact solution to which we can compare our numerical results.

We write the metric in the Kerr-Schild coordinates; as a result, all the variables are well-behaved at the horizon (horizon penetrating).
We fix the metric for this test and evolve the fluid equations only.

For this test, we evolve the same configuration used by Duez et al.~\cite{doi:10.1103/PhysRevD.72.024028}.
The accretion rate is $\dot{M} = 1$, the sonic radius is at $r=8M$ (where $M$ is the mass of the black hole), and the equation of state obeys a $\Gamma = 4/3$ power law [see Eqs.~(\ref{eq:eos-43-pressure})--(\ref{eq:eos-43-epsilon})].
We freeze the hydro evolution variables at the inner and outer boundaries.
We set the inner boundary radius outside of the horizon at $r=2.8M$ (the horizon is at $r=2M$), and the outer boundary is placed at $r=9M$; the Cartesian grid extends $\pm10M$ along each axis.

We evolve this accretion flow at three different resolution: $64^3$, $96^3$ and $128^3$. 
The initial magnetic field is radial such that $b^2/\rho= 1$ and the solution is stationary.
Reconstruction is performed using WENO5.
We add Kreiss-Oliger dissipation~\cite{Kreiss:1973aa} to the evolution of all conservative variables.
This removes short-wavelength noise that would otherwise interfere with clean convergence.

\begin{figure}
\includegraphics[width=\linewidth]{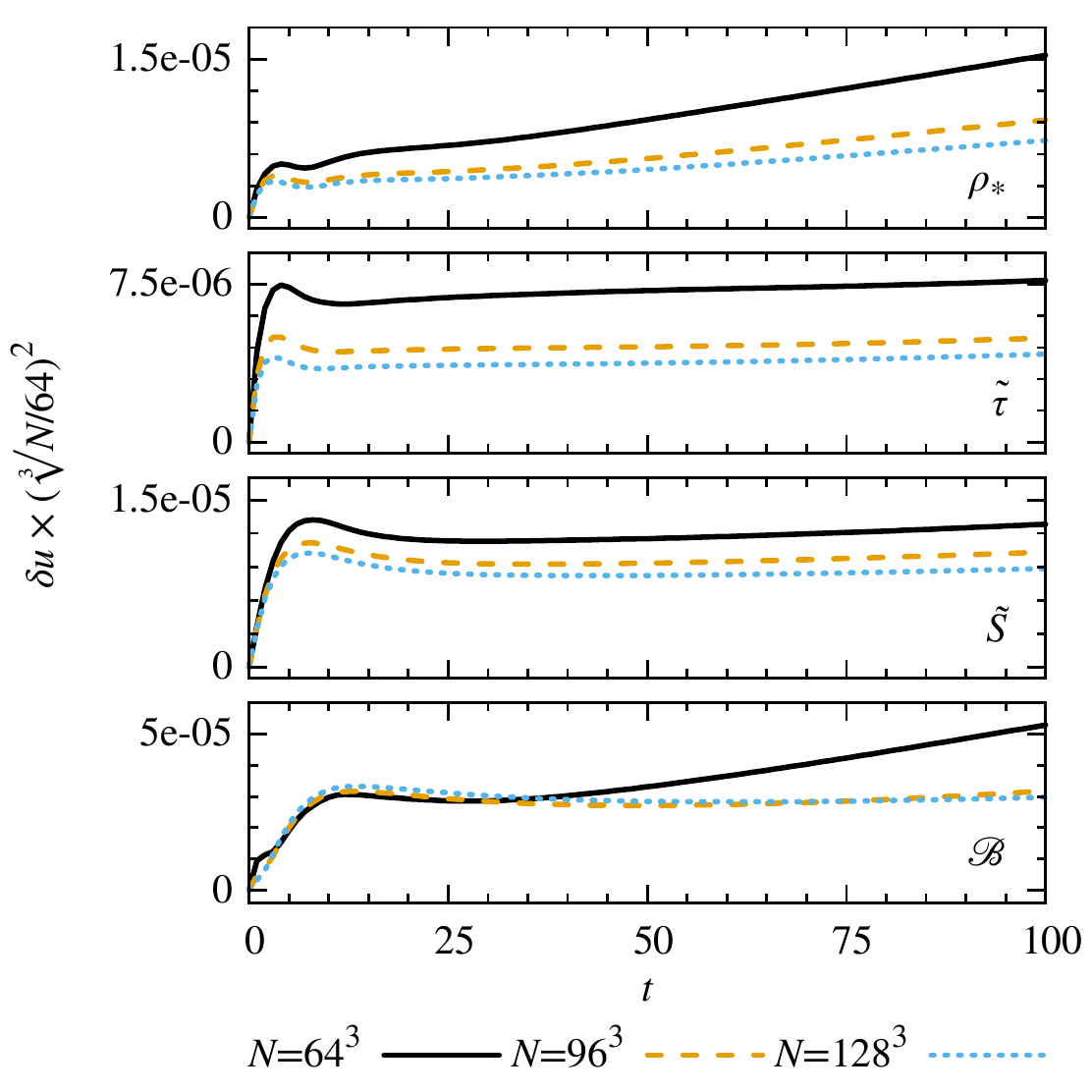}
\caption{\label{fig:bondi-errors}(color online).  Error norm for the Bondi test at three resolutions, rescaled for second-order convergence.}
\end{figure}

We compute the volume $L_2$ norm of the deviation of the conservative variables from their exact Bondi solutions:
\begin{equation}
\delta u = \left(\frac{\int|u-u_\text{exact}|^2\sqrt{\gamma}d^{3}x}{\int\sqrt{\gamma}d^{3}x}\right)^{1/2}.
\label{eq:delta}
\end{equation}
In Fig.~\ref{fig:bondi-errors} we plot the error norm measured by Eq.~(\ref{eq:delta}) for all conservative 
variables after $100M$ of evolution for three different resolutions.
These show that our results are converging at second-order, as expected (and as also observed in previous
studies of this problem, e.g.~\cite{doi:10.1088/0264-9381/24/12/S16, doi:10.1088/0264-9381/31/1/015005}).

\section{\label{sec:zernike-spec}Spectral method for cylinders and spheres}
When evolving the spacetime metric on a spectral grid, we try to adapt the domain decomposition to the geometry of the evolved fields.
This often means using sections of a sphere, in the form of spherical shells or ``cubed spheres.''
In black hole spacetimes, this is sufficient to cover the area surrounding the excised region within the apparent horizon.
However, for neutron star spacetimes, a different approach is taken to cover the center of the star.

Polar and spherical coordinates are singular at the origin, creating difficulties if one tries to use tensor products of one-dimensional function bases.
This same problem exists at the poles of a spherical surface.
Spherical harmonics, $Y^m_l(\theta, \phi)$, provide a clean solution in that case, able to represent smooth functions without artificial boundaries and without severely restricting the timestep allowed by the Courant-Friedrichs-Lewy stability limit~\cite{doi:10.1147/rd.112.0215}.
For the radial ``pole problem,'' Zernike polynomials and their higher dimensional generalizations provide a similar solution.

The use of Zernike polynomials in spectral methods over the unit disk was explored independently by Matsushima \& Marcus~\cite{doi:10.1006/jcph.1995.1171} and by Verkley~\cite{doi:10.1006/jcph.1997.5747}.  Notation varies throughout the literature, so we summarize ours here:

Denote an orthonormal azimuthal (Fourier) basis as
\begin{equation}
  F_m(\phi) \equiv
  \begin{cases}
    \frac{1}{\sqrt{2\pi}} & m=0 \\
    \frac{1}{\sqrt{\pi}} e^{i m \phi} & m>0
  \end{cases} \,.
\end{equation}
Then an arbitrary smooth function $f(\varpi, \phi)$ over the unit disk can be decomposed into its Fourier coefficients $f_m(\varpi)$:
\begin{equation}
  f(\varpi, \phi) = \Re \sum_{m=0}^{m_\text{max}} f_m(\varpi) F_m(\phi) \,,
\end{equation}
where $m_\text{max} = \lfloor N_\phi/2 \rfloor$, $N_\phi$ being the number of azimuthal collocation points.
(Note that if $N_\phi$ is odd, the highest mode will lack a sine component.)

These Fourier coefficients can be further decomposed into a radial sub-basis $R^m_n(\varpi)$, composed of one-sided Jacobi polynomials multiplied by $\varpi^m$:
\begin{equation}
  R^m_n(\varpi) \equiv
    \sqrt{2n+2} \varpi^m P_{(n-m)/2}^{(0,m)}(2 \varpi^2 - 1) \,,
\end{equation}
where $P_k^{(\alpha, \beta)}(x)$ represents the Jacobi polynomial of degree $k$.
In this notation, the radial functions are only defined for $n \geq m$, $2|(n-m)$.
For smooth functions, the $f_m(\varpi)$ satisfy the pole condition: $f_m(\varpi) \rightarrow \varpi^m$ as $\varpi \rightarrow 0$.
This basis manifestly respects that condition.

The Zernike polynomials are then defined as
\begin{equation}
  Z^n_m(\varpi,\phi) \equiv R^m_n(\varpi) F_m(\phi) \,.
\end{equation}
They form an orthonormal basis for smooth functions over the unit disk:
\begin{equation}
  f(\varpi,\phi) = \sum_{m=0}^{m_\text{max}}
    \sum_{\substack{
      n=m \\
      n += 2
    }}^{n_\text{max}} f_{nm} Z^n_m(\varpi, \phi) \,,
\end{equation}
where $n_\text{max}=2N_\varpi -1$, $N_\varpi$ being the number of radial collocation points.
Note that if Gauss-Radau quadrature is used (placing collocation points on the outer boundary of the disk), then the highest-order radial basis functions should be normalized with respect to the quadrature rule (rather than analytically) or else omitted entirely.
Specifications for the quadrature nodes and weights can be found in the references.

As mentioned by Livermore et al.~\cite{doi:10.1016/j.jcp.2007.08.026}, this can be generalized to filled spheres.
In that case, a function $f(r, \theta, \phi)$ is decomposed into $f_{nlm}$ such that
\begin{equation}
  f(r,\theta,\phi) = \sum_{m=-m_\text{max}}^{m_\text{max}} \sum_{l=|m|}^{l_\text{max}}
    \sum_{\substack{
      n=l \\
      n += 2
    }}^{n_\text{max}} f_{nlm} R^l_n(r) Y^m_l(\theta, \phi) \,,
\end{equation}
where now $R^l_n(r)$ is given by
\begin{equation}
  R^l_n(r) = \sqrt{2n+3} r^l P_{(n-l)/2}^{(0,l+1/2)}(2 r^2 - 1) \,,
\end{equation}
which corresponds to an integration weight of $r^2$ instead of $\varpi$.
Here, $Y_l^m(\theta, \phi)$ are the spherical harmonics, and $l_\text{max} = N_\theta - 1$ for $N_\theta$ latitudinal collocation points.

Spectral methods can be susceptible to aliasing instabilities when, for instance, non-linear interactions allow the creation of higher spectral modes through the mixing of lower ones.
Appropriate filtering of the solution is therefore required for stable evolutions~\cite{doi:10.1103/PhysRevD.80.124010}.
When using cylindrical and spherical domains in \textsc{SpEC}, we have found filtering to be unnecessary in the radial direction.
Filtering in angular directions, meanwhile, is performed as for spherical shells~\cite{doi:10.1103/PhysRevD.71.064020}.

\section{\label{sec:zernike}Measuring power in azimuthal modes}
\subsection{Preliminaries}
Consider a function space spanned by a set of $N$ basis functions $\phi_n(x)$ that are orthonormal with respect to a weight function $w(x)$.
That is,
\begin{equation}
\int \phi_m(x) \phi_n(x) w(x) dx = \delta_{mn} \,.
\end{equation}
Further, assume the existence of a quadrature rule on a set of $N$ collocation points $x_i$ that is exact for all products of two functions in this space weighted by $w(x)$.  In other words,
\begin{equation}
\sum_{i=0}^{N-1} \phi_m(x_i) \phi_n(x_i) w_i = \delta_{mn} \,,
\end{equation}
where $w_i$ are the quadrature weights.  Note that Gaussian quadrature meets this criterion for polynomial bases.

Let $f(x)$ be a member of this space, which we write as a linear combination of the basis functions:
\begin{equation}
f(x) = \sum_{n=0}^{N-1} f_n \phi_n(x) \,,
\end{equation}
where the spectral coefficients $f_n$ can be computed via
\begin{equation}
f_n = \int f(x) \phi_n(x) w(x) dx = \sum_{i=0}^{N-1} f(x_i) \phi_n(x_i) w_i \,.
\end{equation}

There exists a unique set of cardinal function $C_i(x)$ in this space with the property that
\begin{equation}
f(x) = \sum_{i=0}^{N-1} f(x_i) C_i(x) \,,
\end{equation}
which we can solve for as follows: First, expand each $C_i(x)$ into its spectral coefficients $c_{n,i}$.  Then we have
\[
f(x) = \sum_{i=0}^{N-1} f(x_i) C_i(x) = \sum_{i=0}^{N-1} f(x_i) \sum_{n=0}^{N-1} c_{n,i} \phi_n(x) \,,
\]
which implies that
\[
\sum_{n=0}^{N-1} f_n \phi_n(x) = \sum_{n=0}^{N-1} \left( \sum_{i=0}^{N-1} f(x_i) c_{n,i} \right) \phi_n(x) \,,
\]
and thus that
\[
f_n = \sum_{i=0}^{N-1} f(x_i) \phi_n(x_i) w_i = \sum_{i=0}^{N-1} f(x_i) c_{n,i} \,.
\]
This means that
\[
c_{n,i} = \phi_n(x_i) w_i \,,
\]
and therefore
\begin{equation}
C_i(x) = w_i \sum_{n=0}^{N-1} \phi_n(x_i) \phi_n(x) \,.
\end{equation}

Observe that the cardinal functions obey the property
\begin{equation}
C_i(x_j) = \delta_{ij}
\end{equation}
and are orthogonal to one another with norm $\sqrt{w_i}$:
\begin{equation}
\int C_i(x) C_j(x) w(x) dx = w_i \delta_{ij} \,.
\end{equation}
Thus, the functions $\tilde{C}_i(x) \equiv C_i(x)/\sqrt{w_i}$ form another orthonormal basis for the space.  (Note that this also provides a convenient way of computing the quadrature weights via $1/w_i = \sum_n \phi_n^2(x_i)$.)

\subsection{Azimuthal power}
Within the space of smooth functions defined in a cylindrical volume, consider the subspace spanned by a finite number of orthonormal basis functions of the form $P_l(z) Z^n_m(\varpi, \phi)$, where $P_l(z)$ is a basis for functions on a finite interval (such as Legendre polynomials) and $Z^n_m(\varpi, \phi) = R^m_n(\varpi) F_m(\phi)$ are the Zernike polynomials (see Appendix~\ref{sec:zernike-spec} for notation).
Any function $f$ in this subspace can be decomposed into spectral coefficients $f_{lmn}$.  The amount of power in a given azimuthal mode $m$ is defined to be
\begin{equation}
  \label{eq:mpower}
P_m[f] = \sum_l \sum_n |f_{lmn}|^2 \,.
\end{equation}
One approach to computing this power for an arbitrary $f$ is to compute each $f_{lmn}$ by integrating $f(z, \phi, r)$ against the corresponding product of basis functions.  If $f$ is band-limited and the integration is of sufficiently high order, this will produce the exact result.  Alternatively, $f$ can be integrated against the set of cardinal functions along $z$ and $r$.  Here we show the equivalence of this nodal approach to the aforementioned modal one.

Let us denote our nodal power measurement by $Q_m[f]$:
\begin{equation}
Q_m[f] \equiv \sum_{i,j} \left| \iiint dz d\phi \varpi d\varpi f(z, \phi, \varpi) \tilde{C}_i(z) \tilde{C}^m_j(\varpi) F_m(\phi) \right|^2 \,;
\end{equation}
here, $\tilde{C}_i(z)$ are the normalized cardinal functions associated with $P_l(z)$ and $\tilde{C}^m_j(\varpi)$ are the normalized cardinal functions associated with $R^m_n(\varpi)$.
Expanding those cardinal functions in terms of their associated basis functions yields
\begin{widetext}
\begin{equation}
Q_m[f] = \sum_{i,j} \left| \iiint dz d\phi \varpi d\varpi f(z, \phi, \varpi) \left( \sqrt{w^P_i} \sum_l P_l(z_i) P_l(z) \right) \left( \sqrt{w^R_j} \sum_n R^m_n(\varpi_j) R^m_n(\varpi) \right) F_m(\phi) \right|^2 \,.
\end{equation}
\end{widetext}
The presence of the weights suggests that the outer sums can be interpreted as integrals (note that the corresponding integrands are products of two basis functions and therefore exactly integrable by quadrature).  And since the basis functions are orthonormal, the integral of a product of sums is equal to a sum of products.  This simplifies the above expression to
\begin{equation}
Q_m[f] = \sum_{l, n} \left| \iiint dz d\phi \varpi d\varpi f(z, \phi, \varpi) P_l(z) R^m_n(\varpi) F_m(\phi) \right|^2 \,.
\end{equation}
But the integral above is merely the projection of $f$ onto the basis function indexed by $l,m,n$; thus
\begin{equation}
Q_m[f] = \sum_{l,n} |f_{lmn}|^2 = P_m[f] \,.
\end{equation}
This gives us two formally equivalent ways to measure the azimuthal power in $f$: one involving projections onto the modal basis, the other projecting onto the nodal (cardinal) basis.  The latter matches an intuitive approach to avoiding the problem of power cancellation due to phase changes at different $\varpi$ and $z$.

\subsection{Error floor}

Unfortunately, when performing these integrations on a finite volume domain, the Cartesian nature of the grid results in spurious power in $m=4,8,\ldots$ modes proportional to the error of the integration scheme (these ``ambients grid modes'' are also noted in studies where mode measurement is restricted to rings~\cite{doi:10.1088/0264-9381/24/12/S10, doi:10.1051/0004-6361:20078577}).
If the function does not approach zero at the boundary of the reference cylinder, then this spurious power will be significant because of the ``Lego circle'' approximation to the boundary.

This effect can be mitigated by windowing the data with a smooth function that transitions between one at the center and zero at the boundary. 
We have achieved good results using the window
\begin{equation}
  W(\varpi) = \frac{1}{2} \left\{1 - \tanh\left[\tan\left(\pi\left(\varpi + 1/2\right) \right)\right] \right\} \,.
\end{equation}
The effect of the windowing on the power spectrum can then be undone via a deconvolution (made robust by using a truncated singular value decomposition).  Expressing the convolution of the spectrum as
\begin{equation}
C_{ij} \lambda_j = \lambda^\prime_i \,,
\end{equation}
the elements of $\bm{C}$ are given by
\begin{equation}
C_{ij} = \int W(\varpi) R^m_i(\varpi) R^m_j(\varpi) \varpi d\varpi \,.
\end{equation}

However, if the function being analyzed is entirely contained within the reference cylinder (by making its radius larger than that of the star, for instance), then this windowing technique offers minimal improvement to the error floor.
Additionally, for our setup, evolved data exhibits $100\times$ more spurious power than initial data.
The net result is that, at our resolution, $m=4$ perturbations can only be measured if they are larger than \num{e-5} relative to the background.
The act of windowing does make this procedure more robust, however, should the data expand beyond the chosen reference cylinder.

\bibliography{localReferences}

\end{document}